\documentclass{aa}

\usepackage{enumitem}
\usepackage{graphicx}
\usepackage{txfonts}
\usepackage{booktabs}

\usepackage[colorlinks=true,allcolors=blue]{hyperref}
\usepackage{float}
\usepackage{placeins}

\begin{document}

   \title{First upper limits on the 21 cm signal power spectrum from cosmic dawn from one night of observations with NenuFAR}
   \titlerunning{First upper limits on the 21 cm signal power spectrum from cosmic dawn with NenuFAR}

   \author{S. Munshi\inst{1}
          \and
          F. G. Mertens\inst{2,1}
          \and
          L. V. E. Koopmans\inst{1}
          \and
          A. R. Offringa\inst{3,1}
          \and
          B. Semelin\inst{2}
          \and
          D. Aubert\inst{4}
          \and
          R. Barkana\inst{5,6,7}
          \and
          A. Bracco\inst{8}
          \and
          S. A. Brackenhoff\inst{1}
          \and
          B. Cecconi\inst{9,10}
          \and
          E. Ceccotti\inst{1}
          \and
          S. Corbel\inst{10,14}
          \and
          A. Fialkov\inst{11,12}
          \and
          B. K. Gehlot\inst{1}
          \and
          R. Ghara\inst{13}
          \and
          J. N. Girard\inst{9}
          \and
          J. M. Grie{\ss}meier\inst{15,10}
          \and
          C. H\"{o}fer\inst{1}
          \and
          I. Hothi\inst{2,8}
          \and
          R. M\'{e}riot\inst{2}
          \and
          M. Mevius\inst{3}
          \and
          P. Ocvirk\inst{4}
          \and
          A. K. Shaw\inst{13}
          \and
          G. Theureau\inst{15,10,16}
          \and
          S. Yatawatta\inst{3}
          \and
          P. Zarka\inst{9,10}
          \and
          S. Zaroubi\inst{13,1}
          }

   \institute{Kapteyn Astronomical Institute, University of Groningen, P.O. Box 800, 9700 AV Groningen, The Netherlands
         \and
    LERMA, Observatoire de Paris, Universit\'{e} PSL, CNRS, Sorbonne Universit\'{e}, F-75014 Paris, France
        \and
    ASTRON, PO Box 2, 7990 AA Dwingeloo, The Netherlands
        \and
    Universit\'e de Strasbourg, CNRS, Observatoire astronomique de Strasbourg (ObAS), 11 rue de l’Universit\'e, F-67000 Strasbourg, France
        \and
    School of Physics and Astronomy, Tel Aviv University, Tel Aviv, 69978, Israel
        \and
    Institute for Advanced Study, 1 Einstein Drive, Princeton, New Jersey 08540, USA
        \and
    Department of Astronomy and Astrophysics, University of California, Santa Cruz, CA 95064, USA
        \and
    Laboratoire de Physique de l'Ecole Normale Sup\'erieure, ENS, Universit\'e PSL, CNRS, Sorbonne Universit\'e, Universit\'e de Paris, F-75005 Paris, France
        \and
    LESIA, Observatoire de Paris, CNRS, PSL, Sorbonne Université, Université Paris Cité, Meudon, France
        \and
    ORN, Observatoire de Paris, CNRS, PSL, Université d'Orléans, Nançay, France
        \and
    Kavli Institute for Cosmology, Madingley Road, Cambridge CB3 0HA, UK
        \and
    Institute of Astronomy, University of Cambridge, Madingley Road, Cambridge CB3 0HA, UK
        \and
    ARCO (Astrophysics Research Center), Department of Natural Sciences, The Open University of Israel, 1 University Road, PO Box 808, Ra’anana 4353701, Israel
        \and
    AIM, CEA, CNRS, Université Paris Cité, Université Paris-Saclay, F-91191 Gif-sur-Yvette, France
        \and
    LPC2E, Université d'Orléans, CNRS, F-45071 Orléans, France
        \and
    LUTh, Observatoire de Paris, Université PSL, Université de Paris Cité, CNRS, F-92190 Meudon, France
             }

   \date{Received 19 October 2023 / Accepted 3 November 2023}
 
  \abstract
   {The redshifted 21 cm signal from neutral hydrogen is a direct probe of the physics of the early universe and has been an important science driver of many present and upcoming radio interferometers. In this study we use a single night of observations with the New Extension in Nan\c cay Upgrading LOFAR (NenuFAR) to place upper limits on the 21 cm power spectrum from cosmic dawn at a redshift of $z$ = 20.3. NenuFAR is a new low-frequency radio interferometer, operating in the 10--85~MHz frequency range, currently under construction at the Nan\c cay Radio Observatory in France. It is a phased array instrument with a very dense $u \varv$ coverage at short baselines, making it one of the most sensitive instruments for 21 cm cosmology analyses at these frequencies. Our analysis adopts the foreground subtraction approach, in which sky sources are modeled and subtracted through calibration and residual foregrounds are subsequently removed using Gaussian process regression. The final power spectra are constructed from the gridded residual data cubes in the $u \varv$ plane. Signal injection tests are performed at each step of the analysis pipeline, the relevant pipeline settings are optimized to ensure minimal signal loss, and any signal suppression is accounted for through a bias correction on our final upper limits. We obtain a best 2$\sigma$ upper limit of $2.4\times 10^7$ $\text{mK}^{2}$ at $z$ = 20.3 and $k$ = 0.041 $h\,\text{cMpc}^{-1}$. We see a strong excess power in the data, making our upper limits two orders of magnitude higher than the thermal noise limit. We investigate the origin and nature of this excess power and discuss further improvements to the analysis pipeline that can potentially mitigate it and consequently allow us to reach thermal noise sensitivity when multiple nights of observations are processed in the future.}


   \maketitle
%________________________________________________________________
\section{Introduction}\label{sec:introduction}
The redshifted 21 cm signal arising out of the hyperfine transition of neutral hydrogen (HI) is a very sensitive probe of the astrophysical processes active during the early stages of cosmic evolution \citep{madau199721,shaver1999can}. Cosmic dawn is a particularly interesting epoch in the early universe for which the 21 cm signal can potentially give us rich insights. This is the period when the gas aggregated in dark-matter halos became dense enough to undergo gravitational collapse and form the first luminous objects in the universe. The two main processes believed to dominate during cosmic dawn are Lyman-$\alpha$ coupling and X-ray heating. The Lyman-$\alpha$ radiation emitted by the first sources couples the spin temperature of HI to the kinetic temperature of the intergalactic medium (IGM) through the Wouthuysen-Field effect \citep{wouthuysen1952excitation,field1958excitation}. Due to large spatial fluctuations in the distribution of early galaxies, this Lyman-$\alpha$ coupling is believed to have occurred inhomogeneously, leading to observable 21 cm signal intensity fluctuations \citep{barkana2005method}. At later stages, X-ray radiation by stellar remnants heats the IGM \citep{venkatesan2001heating,chen2004spin,pritchard200721}, and this in turn increases the spin temperature of the 21 cm transition, which is coupled to the IGM temperature. Since ionization can still be largely neglected at the high redshifts of interest, these physical processes guide the magnitude and spatial distribution of the brightness temperature of the 21 cm signal against the cosmic microwave background radiation (CMBR) during this era. Radio-wave observations by interferometric arrays targeting the redshifts corresponding to cosmic dawn can directly probe the fluctuations in the three-dimensional brightness temperature distribution of the 21 cm signal during this epoch. Thus, these observations have the immense potential of providing us with crucial constraints on the properties of the IGM and early star formation histories \citep[e.g.,][]{furlanetto2006cosmology,pritchard201221,fialkov2014rich,mebane2020effects,2020MNRAS.499.5993R,2022MNRAS.516..841G,hibbard2022constraining,munoz2022impact,monsalve2019results,adams2023improved,bevins2023joint}.

The prospect of probing cosmic dawn and the subsequent epoch of reionization (EoR) has led to many interferometric experiments devoting significant observing time to 21 cm cosmology programs. Most of these experiments have focused on the EoR, and over the years increasingly stringent upper limits on the redshifted 21 cm signal power spectrum from the EoR have been set by the GMRT\footnote{Giant Metrewave Radio Telescope} \citep{paciga2013simulation}, PAPER\footnote{Precision Array to Probe EoR} \citep{kolopanis2019simplified}, MWA\footnote{Murchison Widefield Array} \citep{barry2019improving,li2019first,trott2020deep}, LOFAR\footnote{Low-Frequency Array} \citep{patil2017upper,mertens2020improved}, and HERA\footnote{Hydrogen Epoch of Reionization Array} \citep{abdurashidova2022first}. Many current and upcoming interferometric experiments are also trying to detect or place limits on the 21 cm signal from cosmic dawn, with some of them having already set upper limits on the 21 cm power spectrum at $z$ $>$ 15. The first attempt at measuring the 21 cm power spectrum during cosmic dawn was by \citet{ewall2016first}, who used 6~h of MWA data to set upper limits of $\Delta_{21}^{2} < 8.3 \times 10^7 \text{mK}^2$ at $z$ = 15.35 and $k$ = 0.21 $h\text{Mpc}^{-1}$, and $\Delta_{21}^{2} < 2.7 \times 10^8 \text{mK}^2$ at $z$ = 17.05 and $k$ = 0.22 $h\text{Mpc}^{-1}$. Later, \citet{yoshiura2021new} used 5.5 h of MWA data to set an upper limit of $\Delta_{21}^{2} < 6.3\times 10^{6} \text{mK}^2$ at $k$ = 0.14 $h\, \text{Mpc}^{-1}$ and $z$ = 15.2. The OVRO-LWA\footnote{Owen's Valley Radio Observatory - Long Wavelength Array} was used first by \citet{eastwood201921} to provide upper limits of $\Delta_{21}^{2} < 10^{8} \text{mK}^2$ at $z$ $\approx$ 18.4 and $k$ $\approx$ 0.1 $\text{Mpc}^{-1}$ using 28~h of data and later by \citet{garsden202121}, who reported an upper limit of $\Delta_{21}^{2} < 2 \times 10^{12} \text{mK}^2$ at $z$ = 28 and $k$ = 0.3 $h\text{Mpc}^{-1}$ using 4~h of data. \citet{gehlot2019first} set 2$\sigma$ upper limits for the 21 cm power spectrum at $\Delta_{21}^{2} < 2 \times 10^{8} \text{mK}^2$ at $z$ = 19.8 to 25.2 and $k$ = 0.038 $h\, \text{cMpc}^{-1}$ using 14~h of data obtained using the LOFAR LBA\footnote{Low Band Antenna} system. The ACE\footnote{AARTFAAC Cosmic Explorer} program \citep{gehlot2020aartfaac}, which uses the LOFAR AARTFAAC\footnote{Amsterdam ASTRON Radio Transients Facility And Analysis Center}, placed 21 cm power spectrum upper limits of $\Delta_{21}^{2} < 5 \times 10^{7} \text{mK}^2$ at $z$ = 17.9 to 18.6 and $k$ = 0.144 $h\, \text{cMpc}^{-1}$ using 2~h of data.  Second-generation experiments such as HERA \citep{deboer2017hydrogen} and SKA-Low\footnote{Square Kilometer Array - Low} \citep{koopmans2015cosmic} are expected to have higher sensitivities in these redshifts, and SKA-Low has the potential to detect the 21 cm signal from cosmic dawn in tomographic images.

In a complementary approach, several single dipole experiments have attempted to detect the sky-averaged (global) 21 cm signal at high redshifts. Among these, the EDGES\footnote{Experiment to Detect the Global EoR Signature}\citep{bowman2008toward} and SARAS\footnote{Shaped Antenna measurement of the background RAdio Spectrum}\citep{singh2017first,subrahmanyan2021saras} are the only ones currently in the sensitivity range of a detection. In 2018, the EDGES collaboration reported an absorption trough in the global 21 cm signal at the redshifts corresponding to cosmic dawn \citep{bowman2018absorption}. However, the feature detected by EDGES is unusually deep and flat \citep{hills2018concerns, bowman2018reply}, and it becomes necessary to introduce "exotic" nonstandard models of the 21 cm signal to explain it if it is cosmological. One such approach to modeling the EDGES signal involves supercooling of gas by scattering off cold dark-matter particles \citep[e.g.,][]{barkana2018possible,fialkov2018constraining,berlin2018severely,munoz2018small,liu2019reviving}. Another alternative class of models suggests that the background against which we observe the 21 cm signal could have a component in addition to the CMBR, thus explaining the unusual depth of the trough \citep[e.g.,][]{feng2018enhanced,ewall2018modeling,dowell2018radio,fialkov2019signature}. These models, in turn, predict stronger fluctuations in the 21 cm signal from cosmic dawn, implying that the 21 cm power spectrum could potentially be easier to detect. However, as of now, no independent confirmation of the EDGES result has been obtained by other experiments, and a recent measurement by the SARAS team claims to disprove the EDGES detection with a 95\% confidence level \citep{singh2022detection}.

In this work we present the first upper limits on the 21 cm signal power spectrum from cosmic dawn using NenuFAR\footnote{New Extension in Nan\c cay Upgrading LOFAR} \citep{zarka2012lss}, a radio interferometer nearing completion at the Nan\c cay Radio Observatory in France. NenuFAR is a low-frequency phased array instrument that will have 1976 receiving antennas, with a very dense core and a large collecting area, making it extremely sensitive to the large scales necessary for constraining the 21 cm signal \citep{joseph201221}. It is one of the most sensitive instruments for cosmic dawn observations and is currently the most sensitive instrument below 50~MHz. In this study we performed an end-to-end analysis of a single night of observations made with an incomplete array composed of 79 active interferometric elements. We used an analysis method that adopts many of the strategies that have been developed and improved over the years during the evolution of the LOFAR EoR analysis pipeline \citep{chapman2012foreground,chapman2013scale,patil2014constraining,patil2016systematic,patil2017upper,asad2015polarization,asad2016polarization,asad2018polarization,mertens2018statistical,gehlot2018wide,gehlot2019first,offringa2019precision,hothi2021comparing,mevius2022numerical,gan2023assessing}.
We demonstrate the performance of the NenuFAR calibration pipeline on a single night of observations made with the current incomplete array and present its first upper limits on the 21 cm signal power spectrum. This pilot analysis also reveals the limitations of the current processing pipeline and helps shape our strategy for overcoming them in the future.

The paper is organized as follows: Section \ref{sec:cosmicdawnksp} introduces the NenuFAR instrument and its Cosmic Dawn Key Science Program (CD KSP). In Sect. \ref{sec:preprocessing} we describe the data acquisition process and the preprocessing steps. The different steps in the data calibration procedure are described in Sect. \ref{sec:calibration}. In Sect. \ref{sec:postflag} we describe an additional step of post-calibration flagging. Section \ref{sec:ps} describes the power spectrum estimation, and Sect. \ref{sec:mlgpr} describes the first application of the machine learning (ML) Gaussian process regression  (GPR) foreground removal technique \citep{mertens2023retrieving} to our data. Section \ref{sec:robustness} summarizes our results and presents the robustness tests that were performed to estimate the level of signal suppression due to the analysis pipeline. In Sect. \ref{sec:discussion} we discuss the limitations and future improvements. A short summary of the main conclusions from this analysis is presented in Sect. \ref{sec:summary}. Throughout this paper, a flat $\Lambda$ cold dark matter cosmology is used, with the cosmological parameters ($\mathrm{H}_0=67.7$, $\Omega_m=0.307$) consistent with results from \textit{Planck} observations \citep{planck2016planck}.

\section{NenuFAR Cosmic Dawn Key Science Program}\label{sec:cosmicdawnksp}
The NenuFAR\footnote{\url{https://nenufar.obs-nancay.fr/en/homepage-en/}} CD KSP\footnote{\url{https://vm-weblerma.obspm.fr/nenufar-cosmic-dawn/}} (KP01, P.I.: L. V. E. Koopmans, B. Semelin and F. G. Mertens) is one of the programs that started in the early scientific phase of NenuFAR, which aims to detect or place stringent limits on the 21 cm power spectrum from cosmic dawn in the redshift range of $z$ = 15--31 \citep{2021sf2a.conf..211M}. The project has already been allocated 1267 hours of observations, which started in September 2020 and are still going on with an array that is progressively more sensitive as more core and remote stations are being added. With NenuFAR's sensitivity, detection of the 21 cm signal predicted by EDGES-inspired exotic models is possible within 100 hours of integration. Moreover, with 1000 hours of integration, NenuFAR is expected to reach a thermal noise sensitivity comparable to the levels of the signal predicted by more standard models \citep{mesinger201121cmfast, Murray2020, semelin2023accurate}.

\subsection{The instrument}
\begin{figure}
        \includegraphics[width=\hsize]{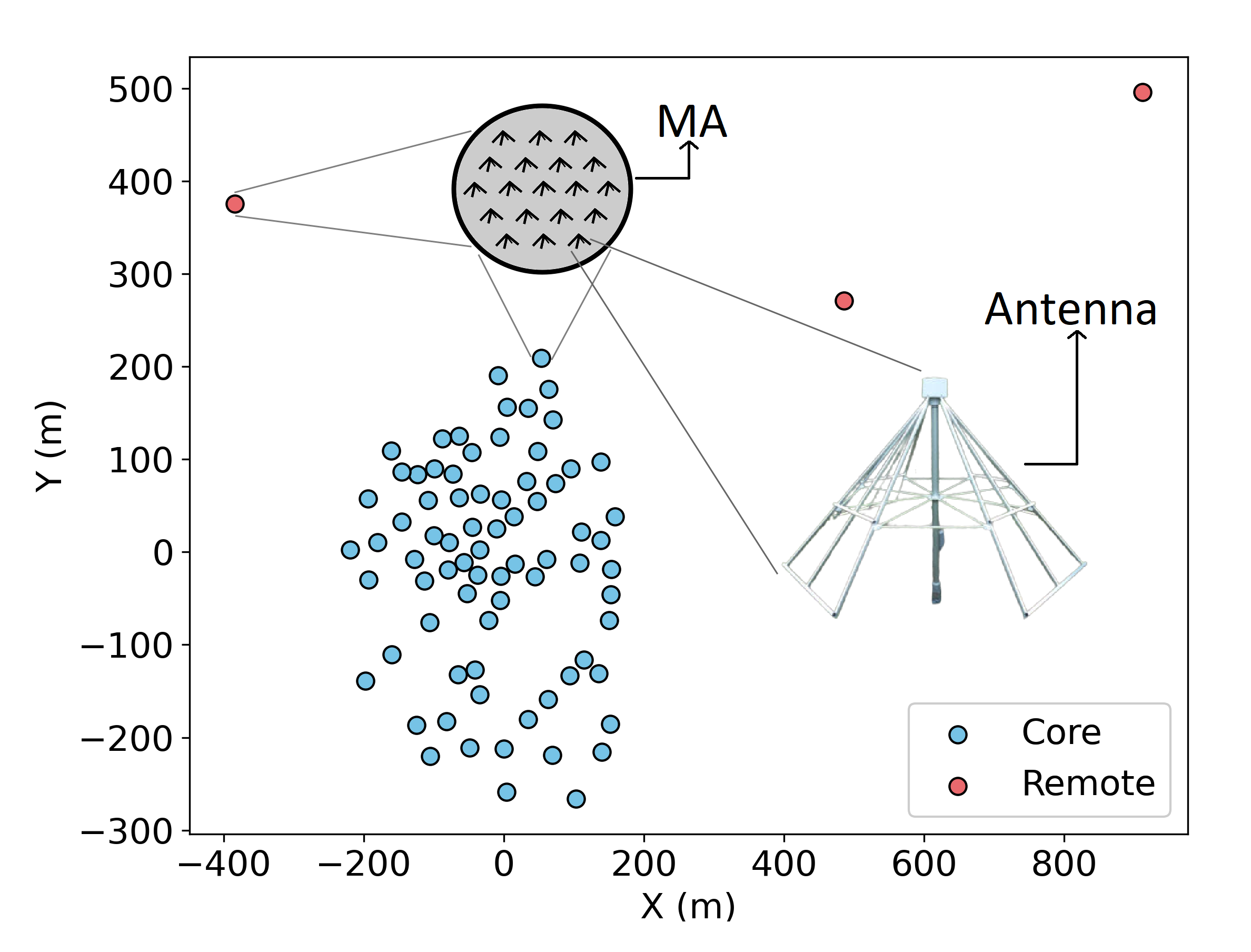}
    \caption{Schematic representation of the NenuFAR array configuration used in the observation.}
    \label{fig:antenna_array}
\end{figure}

NenuFAR \citep{zarka2015nenufar, zarka2020low} is a large low-frequency radio interferometer. The primary receiving antennas for NenuFAR are crossed inverted V-shaped dual polarization dipoles, similar to the antenna elements employed by the LWA \citep{hicks2012wide}, but equipped with an original custom-made low noise amplifier \citep{charrier2007design}. In total, 19 such antennas, arranged in a hexagonal array, form a single mini-array (MA), whose station beam is analog steerable in different directions in the sky. The hexagonal MA has a six-fold symmetry, resulting in considerable grating lobes in an individual MA primary beam. In order to reduce the level of these grating lobes, the MAs are rotated with respect to each other by angles that are multiples of 10 degrees, with only 6 such non-redundant orientations being necessary due to their hexagonal symmetry. At completion, NenuFAR will have 96 MAs distributed in a 400~m diameter core and 8 remote MAs located at distances up to a few kilometers from the center. Figure \ref{fig:antenna_array} presents a schematic representation of the NenuFAR configuration used for obtaining the data used in this analysis. The dense core of NenuFAR with a large number of short baselines makes it extremely sensitive to large scales and especially well suited to constraining the 21 cm signal. The longer baselines involving the remote stations, combined with pointing observations and Earth rotation synthesis, yield a good $u\varv$ coverage out to several kilometers (3.5--5~km at completion, currently 1.4~km) and thus enable efficient point source foreground modeling and subtraction.

\begin{table}[t]
\caption{Observation specifications for the raw data (L0) used in the analysis.}
\label{tab:obsparam}
\centering
\begin{tabular}{@{}ll@{}}
\toprule
Parameter                   & Value            \\ \midrule
Core MAs                    & 76               \\
Remote MAs                  & 3                \\
Spectral Window             & 61.1 - 72.2 MHz \\
Bandwidth                   & 11.1 MHz         \\
Spectral Resolution         & 3.05 kHz         \\
Sub-bands                   & 59               \\
Channels per sub-band           & 64               \\
Maximum Baseline            & 1.46 km          \\
Target                      & NCP              \\
Calibrator                  & Cas A            \\
Target Start Time (UTC)     & 17:33 12/12/2021 \\
Target Duration             & 11.4 hour        \\
Calibrator Start Time (UTC) & 17:00 12/12/2021 \\
Calibrator Duration         & 30 min        \\
Time Resolution             & 1 s         \\ \bottomrule
\end{tabular}
\end{table}

\begin{table*}[t]
\caption{Calibration parameters used in {\tt DDECal} at different stages of the analysis.}
\label{tab:calparam}
\centering
\begin{tabular}{@{}lllll@{}}
\toprule
Parameter  & Bandpass calibration     & A-team subtraction   & 3C subtraction       & NCP subtraction      \\ \midrule
Input Data Level       & L1     & L2     & L2     & L3     \\
Output Data Level       & L2     & L2     & L3     & L4     \\
Mode       & Diagonal     & Diagonal     & Diagonal     & Diagonal     \\
Time Interval  & 4 s         & 2 min        & 4 min        & 8 min        \\
Frequency Interval  & 15.3 kHz     & 195.3 kHz    & 195.3 kHz    & 195.3 kHz    \\
Directions & Cas A        & Four A-team sources and NCP   & Seven 3C sources and NCP       & 7 Clusters   \\
Smoothness constraint  & None        & 2 MHz        & 2 MHz        & 6 MHz        \\
$u\varv$ range   & >15$\lambda$ & >20$\lambda$ & >40$\lambda$ & >40$\lambda$ \\ \bottomrule
\end{tabular}
\tablefoot{L1 data have a 15.26~kHz frequency resolution while L2, L3 and L4 data have 61.03~kHz frequency resolution. The time resolution for L1, L2, L3 and L4 data levels is 4 s.}
\end{table*}

\subsection{Observations and data selection}
The primary target field for the NenuFAR CD KSP is the north celestial pole (NCP). The NCP is a particularly favorable field since it is visible at night throughout the year, allowing long observations on winter nights. The NCP is in a fixed direction, thus requiring no beam or phase tracking with time. Moreover, extensive sky models of the NCP, made using deep integrations with LOFAR, are readily available as a byproduct of the LOFAR EoR Key Science Project (KSP) and these aid the analysis and calibration efforts. The NCP observations with NenuFAR are preceded or succeeded by a 30 min observation of a calibrator -- Cygnus A (Cyg A) or Cassiopeia A (Cas A), depending on which one of the two sources is at a high enough altitude to yield good quality observations.

Phase I observations of the NCP with NenuFAR started in September 2019 with the primary goal of understanding the instrument and monitoring its performance. Deep integrations of the NCP field commenced with the start of Phase II observations in September 2020 and as of May 2022, 1080 hours of NCP data have been gathered, with the newer observations being carried out with more core and remote stations. In July 2021 the number of core MAs was increased from 56 to 80 and in September 2021, the third remote MA was installed. The night used in this analysis was selected on the basis of considerations related to low ionospheric activity among the observations that were available with three remote MAs active as of May 2022 when this pilot survey was initiated. Calibration solutions for the calibrator (Cyg A or Cas A) observations were obtained for all these nights and the temporal fluctuations in these solutions were used as a proxy for ionospheric activity. The trend in the temporal fluctuations as a function of observation nights was verified to be roughly consistent with the trend in the total electron content values for these nights obtained independently from GPS measurements using {\tt RMextract}\footnote{\url{https://github.com/lofar-astron/RMextract/tree/master}}\citep{2018ascl.soft06024M}. The night of 12-13 December 2021 had mild ionospheric activity and coincidentally relatively low level of flagged data due to radio-frequency interference (RFI). This night was selected for the current analysis.

\begin{figure}
        \includegraphics[width=\hsize]{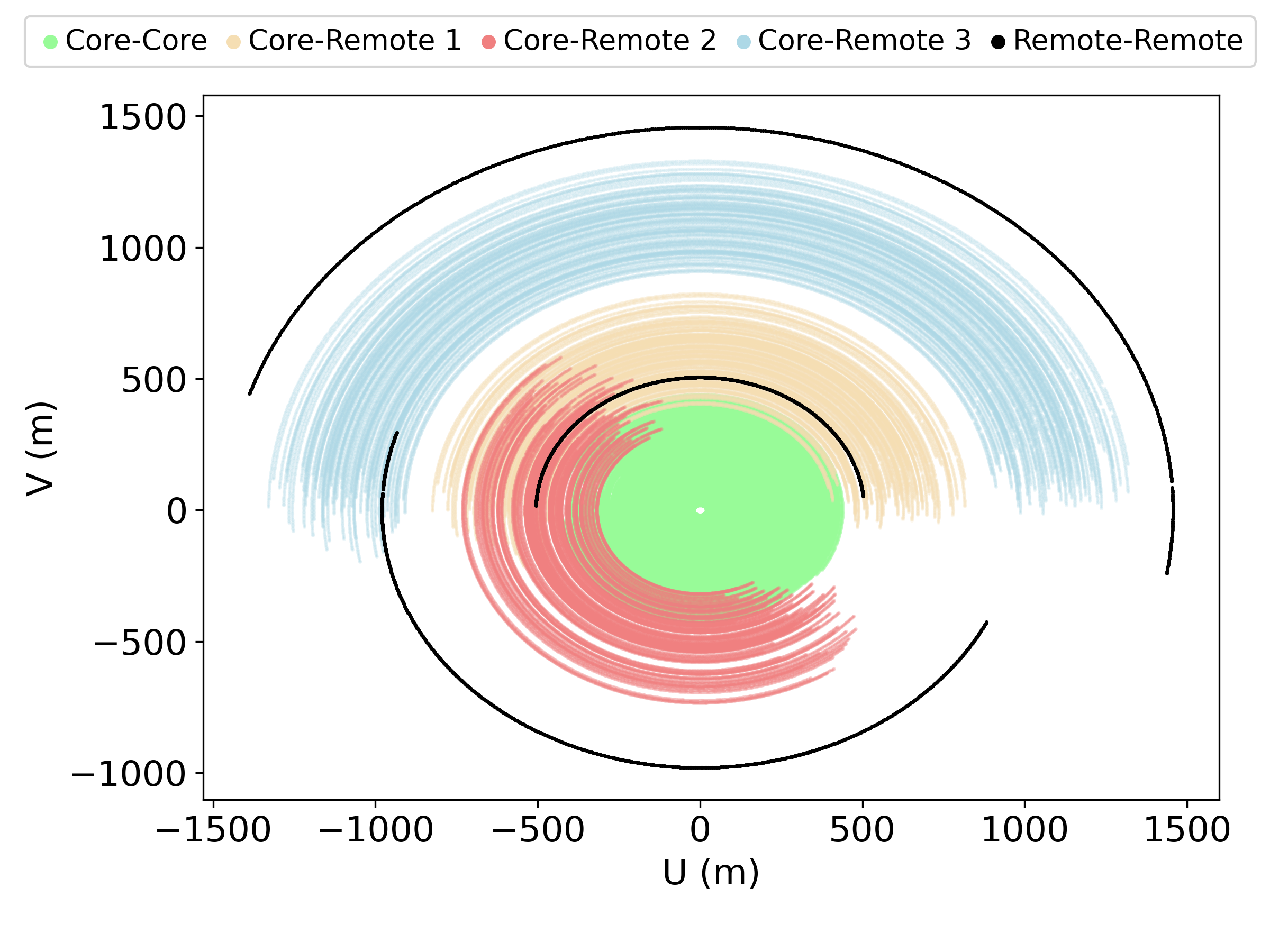}
    \caption{Baseline coverage in the $u \varv$ plane for the NCP observation. The baseline tracks for the core-remote baselines are indicated with different colors for the baselines involving the three remote MAs. For visual clarity, only the U,V points are plotted here, not the -U,-V points. The full coverage including the -U,-V points is much more symmetric.}
    \label{fig:uvplot}
\end{figure}

The 11.4 h NCP observation was preceded by a 30 min observation of Cas A, which was used for bandpass calibration. Though the observations are recorded in 4 spectral windows covering a wide frequency range of 37--85~MHz, we selected a single spectral window ranging from 61 to 72~MHz for this analysis on the basis of superior RFI statistics (Appendix \ref{sec:rfi}). The main observation specifications are summarized in Table \ref{tab:obsparam}. Figure \ref{fig:uvplot} shows the baseline coverage of the observation.

\section{Preprocessing}\label{sec:preprocessing}
Observational data obtained with NenuFAR undergo several routine steps of preliminary processing, which prepare it for further stages of precise calibration and foreground subtraction. These steps are described in this section.

\subsection{Data acquisition}
The NenuFAR correlator is a replica of the COBALT-2 correlator used by LOFAR \citep{broekema2018cobalt} and is called NenuFAR Imager Correlation Kluster Elaborated from LOFAR's (NICKEL). During a NenuFAR observation, the two linear polarization components of the signal from all MAs reach the receiver where they are sampled at a frequency of 200~MHz, followed by channelization into 512 sub-bands of 195.3~kHz width each using a polyphase filter bank. The data for each second is digitally phased to the target direction and cross correlated across the different polarizations to yield full polarization visibilities in 384 sub-bands with a total bandwidth of 75~MHz. Each sub-band is further channelized into 64 channels with a 3.05~kHz spectral resolution. The raw correlated visibilities are finally exported into a measurement set (MS) format.

\subsection{RFI mitigation, bandpass calibration, data averaging, and compression}
The preprocessing of NenuFAR data is performed using the {\tt nenucal-cd} package\footnote{\url{https://gitlab.com/flomertens/nenucal-cd}}, which in turn uses the various tasks for calibration, flagging and averaging included in the default preprocessing pipeline \citep[${\tt DP}^3$;][]{van2018dppp} used for analysis of LOFAR data. NenuFAR visibility data during preprocessing is available at three data levels: L0, L1, and L2 in the order of decreasing time resolution, frequency resolution, and data volume. The raw visibilities output by the correlator (the L0 data) have a time resolution of 1~s and frequency resolution of 3.05~kHz with 64 channels per sub-band. To capture intermittent and narrowband RFI, the software {\tt AOFlagger} \citep{offringa2012morphological} is used directly at this highest time and frequency resolution to flag the data affected by strong RFI. To avoid the edge effects of the polyphase filter used in channelization, two channels at both ends of the sub-band are flagged, and the remaining data are averaged into 4~s time resolution and 15.26~kHz frequency resolution with 12 frequency channels per sub-band.  This is followed by data compression with {\tt Dysco} \citep{offringa2016compression} to produce the L1 data, leading to the reduction in data volume by a factor of $\approx$ 80 compared to the L0 data. Though this compression introduces some noise into the data, it has been verified through simulations that the compression noise is significantly lower than the expected 21 cm signal. Moreover, the noise is uncorrelated across separate observations taken on different days, and hence averaging the data across different days decreases this compression noise just like thermal noise. The L1 data of the targeted observation of Cas A were calibrated using the software {\tt DDECal} (described in Sect. \ref{sec:calibration}) against a sky model of Cas A consisting of point sources and Gaussians created previously using LOFAR LBA observations.  The parameters used in the bandpass calibration with {\tt DDECal} are listed in Table \ref{tab:calparam}. The bad calibration solutions were flagged based on a threshold-clipping algorithm and the remaining solutions were averaged in time to produce one solution per MA, per channel, and per polarization. These bandpass solutions were then applied to the L1 NCP data. The bandpass calibration sets the approximate amplitude scale of the target visibilities and more importantly, accounts for the cable reflections that can produce rapid frequency fluctuations in the data. The results of bandpass calibration for a few example MAs are shown in Appendix \ref{sec:bandpass}. The bandpass calibration is followed by another step of RFI flagging with {\tt AOFlagger} and averaging to a frequency resolution of 61.03~kHz with three channels per sub-band, to get the L2 data product. These data are now suitable for the next steps of calibration.

\begin{figure*}
        \includegraphics[width=\hsize]{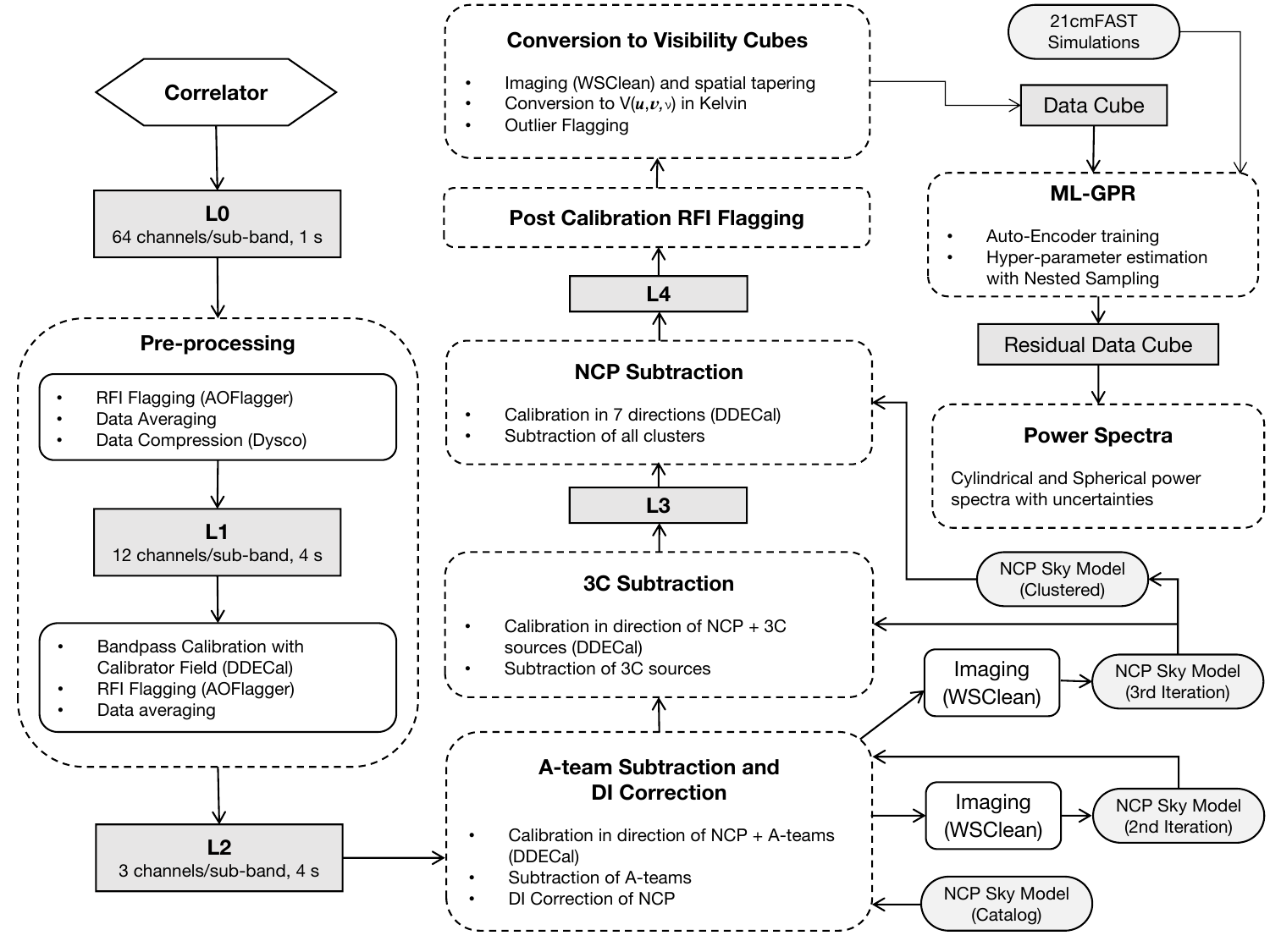}
    \caption{Data processing pipeline with the different steps of preprocessing, calibration, foreground subtraction, and power spectrum estimation. The calibration settings used in the different steps in this pipeline are summarized in Table \ref{tab:calparam}. Note that the signal injection tests performed on the calibration and foreground removal steps are not described in this flowchart (see Fig. \ref{fig:flowchart_injection}).}
    \label{fig:pipeline}
\end{figure*}

\section{Calibration}\label{sec:calibration}
The dataset after the preprocessing was divided into 13 time segments (twelve 52 min segments and one 56 min segment), which were processed in parallel, in 13 computational nodes of the {\tt DAWN} cluster \citep{pandey2020integrated}. This decreased the computation time and enabled us to conduct multiple calibration runs that allowed us to optimize the calibration settings for this dataset.

Calibration of NenuFAR data is done using {\tt DDECal}, a calibration software that is part of the ${\tt DP}^3$. {\tt DDECal} employs a directional solving algorithm similar to the scalar algorithm described by \citet{smirnov2015radio}. In addition to the {\tt fulljones} mode in which all four elements of the gain matrix are solved for, {\tt DDECal} can also be used in the {\tt diagonal} mode in which the off-diagonal elements are set to zero. A detailed description of the {\tt fulljones} algorithm of {\tt DDECal} is presented by \citet{gan2023assessing}. {\tt DDECal} also has a feature to solve for spectrally smooth solutions by applying constraints to the solutions, using a Gaussian kernel of a chosen full width at half maximum (FWHM). The calibration parameters that were used in {\tt DDECal} at different stages of the data processing are listed in Table \ref{tab:calparam}.

The frequency smoothing kernel prevents rapid nonphysical frequency fluctuations in the calibration solutions and ensures that the gain solutions do not over-fit the thermal noise and sources that are not part of the incomplete sky model used in calibration, including the 21 cm signal \citep{mouri2019quantifying, mevius2022numerical}. The smoothing kernel FWHM was chosen by performing multiple signal injection tests with different kernel widths to find a compromise between maximizing the accuracy of source subtraction and minimizing the signal loss (described in Sect. \ref{sec:robustness_ddcal}). At every stage of calibration, we applied a baseline cut of at least 20$\lambda$ and only used longer baselines for the calibration. This was done because the shorter baselines are especially sensitive to the diffuse Galactic emission, which is not present in the sky model used for calibration. If these baselines were used in calibration, this additional diffuse emission would affect the gain solutions and could reappear elsewhere in the final image or on different spatial and frequency scales \citep{patil2016systematic}. A similar calibration strategy is used by the LOFAR EoR KSP, but the difference is that LOFAR uses a much higher baseline cut (250$\lambda$) and computes the power spectrum using only the shorter baselines (between $50$ and $250\lambda$) that are not used in calibration. However, this is not feasible with NenuFAR for all the calibration steps since there are not enough long baselines to get reliable calibration solutions for all MAs, and the baselines used for generating power spectra are also used in one of the calibration steps. This creates a risk of signal suppression during this calibration step and this was assessed during signal injection tests (described in Sect. \ref{sec:robustness_ddcal}). Figure \ref{fig:pipeline} summarizes all the different steps in the processing pipeline used in this analysis, except for the signal suppression tests.

\subsection{Bright source subtraction and direction-independent correction}\label{sec:sourcesub}
The sources Cas A, Cyg A, Taurus A (Tau A), and Virgo A (Vir A), collectively referred to as the A-team, are the brightest astronomical radio sources in the northern celestial hemisphere at low radio frequencies. In addition to these, there are several other radio sources in the northern sky with a high enough flux to have a significant impact on the NCP field through their point spread function (PSF) sidelobes. The spectra of most of these sources are steep, which means that they become extremely bright at the low frequencies that we are interested in. The result is that even though some of these sources are far away from the NCP and are attenuated by the primary beam, the PSF sidelobes of these sources leave a considerable imprint on the NCP field. Naturally, precise subtraction of these bright sources and mitigation of their PSF sidelobes is crucial for our analysis.

\subsubsection{A-team subtraction and direction-independent correction}\label{sec:ateamsub}
The NenuFAR primary beam model is not yet a part of {\tt DDECal}. Since the four A-team sources are well away (at least 30 degrees) from the NCP, we do not expect the MA gains in the direction of these sources to be the same as those in the direction of the NCP, mainly because of the primary beam. Hence, as these sources move through the stationary primary beam, the MA gain solutions in the direction of these sources will carry an imprint of the primary beam pattern. It is, therefore, necessary to perform direction-dependent (DD) calibration in order to obtain separate time and frequency-dependent gain calibration solutions for the NCP field and the individual A-team sources.

\begin{figure}
        \includegraphics[width=\hsize]{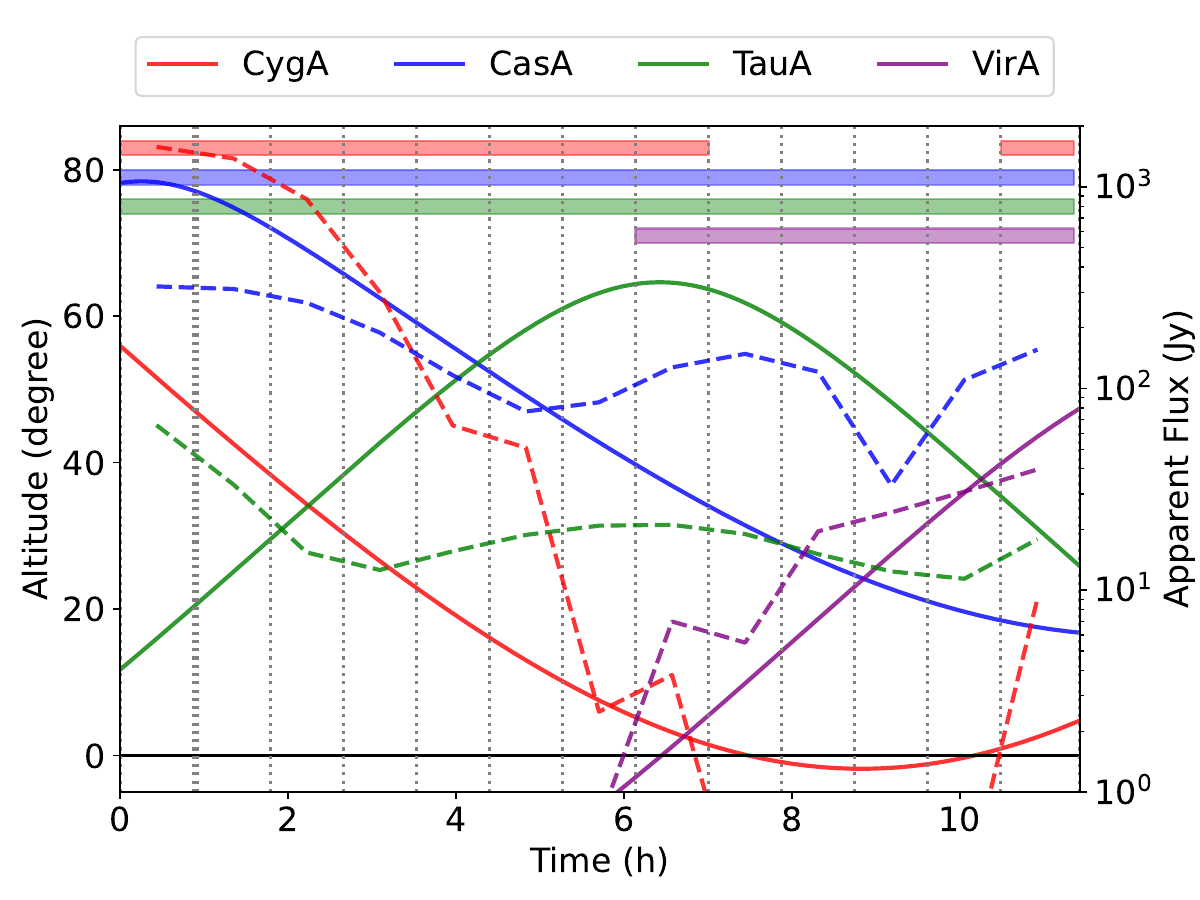}
    \caption{Altitude and beam-attenuated-flux values for the four A-team sources, Cyg A, Cas A, Tau A, and Vir A, that are subtracted in the analysis. The vertical dotted lines demarcate the 13 different segments into which the data are divided.  The solid lines indicate the altitude of the A-team and are to be read using the scale on the left-hand side axis. The dashed lines connect the intrinsic flux values for the A-team attenuated by the averaged preliminary primary beam model value for each data segment and are to be read using the axis on the right-hand side. The four horizontal bands on top indicate the data segments in which each A-team was included in the sky model and hence subtracted through the method described in Sect. \ref{sec:ateamsub}.}
    \label{fig:ateams_altitude}
\end{figure}

We used intrinsic sky models for all the A-team sources, obtained previously using the LOFAR LBA and consisting of point sources (delta-functions) and Gaussians. Not all the A-team sources are above the horizon for all 13 time segments into which we have divided the dataset. So for a given data segment, if a particular A-team source is below the horizon for most of the time, we excluded it from the sky model. This is illustrated in Fig. \ref{fig:ateams_altitude}. For the NCP, we used an intrinsic model obtained from the global sky model\footnote{\url{https://lcs165.lofar.eu/}} that was created using data from The GMRT Sky Survey \citep[TGSS;][]{2017A&A...598A..78I}, NRAO VLA Sky Survey \citep[NVSS;][]{condon1998nrao}, WEsterbork Northern Sky Survey \citep[WENSS;][]{1997A&AS..124..259R}, and VLA Low-frequency Sky Survey \citep[VLSS;][]{cohen2007vla}. For each data segment, the primary beam\footnote{A preliminary model of the NenuFAR primary beam is available from the package {\tt nenupy: }\url{https://github.com/AlanLoh/nenupy/}.}, averaged over all stations, was computed at three equally spaced time points within the segment (corresponding to 1/4, 1/2 and 3/4 of the time duration of the segment) and averaged together to give the station and time averaged beam. The intrinsic sky model components were then attenuated by the beam values at the positions of the components. These "apparent" sky models (as seen by NenuFAR) for the different A-team sources and the NCP were subsequently put in separate sky-model-component clusters to get the final apparent sky model for each data segment. Each such cluster forms a direction for which a separate calibration solution is solved. It should be noted that this kind of attenuation by a simulated beam model accounts for the rotation of the primary beam against the sky on timescales longer than the duration of the time segments (since we used a single sky model for a segment), and up to the accuracy of the preliminary simulated beam model that was used to obtain the apparent sky model used in calibration. The inaccuracy of this preliminary beam model and the spatially averaged effect of the rotation of the beam on timescales as low as the gain solution intervals will be accounted for by the gain calibration solutions. However, since the NenuFAR primary beam model is currently not integrated into {\tt DDECal}, the differences between the actual sky and the apparent sky model within a solution time interval and inside one direction cannot be accounted for and will lead to errors in the calibration.

{\tt DDECal} was used to calibrate the L2 data against the apparent sky model for each time segment. It returns separate calibration solutions for the NCP and each of the A-team sources per MA, polarization, time, and frequency step. Next, the visibilities corresponding to each of the A-team sky models were predicted and the calculated gain solutions were applied to the visibilities. These corrupted predicted visibilities were then subtracted from the data. This removes the A-team from the data if both the source model and the calibration are perfect. The gain solutions in the direction of the NCP were then applied to the data. We note that this is a single calibration solution for the entire NCP field, so this is similar to a direction-independent (DI) correction step if only the NCP field is concerned and this sets the absolute flux scale of the data. This completes the A-team subtraction and DI gain correction on the NCP field using a catalog sky model. The final parameters used for this calibration step were selected after multiple calibration trial runs (described in Appendix \ref{sec:ateamsub_trials}). We find that including all four A-team sources when they are above the horizon, flagging MAs 6, 18, 62, 64, 72, and using {\tt diagonal} mode gives us the optimal results for this dataset.

The next step is to make a sky model of the NCP field using this calibrated NenuFAR data, which will serve as an updated version of the preliminary catalog sky model. This is similar to a self-calibration loop in which the DI gains and the sky model are iteratively updated. First, the A-team subtracted and DI-corrected data were used to make a deconvolved multifrequency synthesis (MFS) image with {\tt WSClean} \citep{offringa2014wsclean}. The parameters used in making the cleaned image are given in Table \ref{tab:imaging} under the column "High Res Clean." The image was made using multi-scale clean \citep{offringa2017optimized} in which both point sources and Gaussians are part of the model. The clean model is given as an output by {\tt WSClean} in the form of a component list, and this provides an updated sky model of the NCP field as observed by NenuFAR. This is an apparent sky model of the NCP because the NenuFAR primary beam, averaged over the observation, is not corrected for in the image. We then used this updated sky model to calibrate the L2 dataset, using the same calibration settings, which gave the best results for the A-team subtraction and DI correction using the catalog sky model. The only difference here is that for each data segment, in place of the intrinsic NCP sky model attenuated by the averaged primary beam, we used this apparent sky model of the NCP. This A-team subtraction and DI correction run provides slightly improved results compared to when the catalog NCP sky model is used, with the final residual image from {\tt WSClean} having 1\% lower rms. The corresponding clean model from a {\tt WSClean} run on the DI corrected data now provides a third iteration model of the NCP field and has a few more sources than the previous sky model. The upper panel of Table \ref{tab:sm} lists the three different sky models used at different stages: the initial catalog sky model from the Global Sky Model ("Initial"), the updated model after A-team subtraction and DI correction with the catalog model ("2nd Iteration") and the third iteration model after A-team subtraction and DI correction with the 2nd Iteration model ("3rd Iteration").

\begin{table}[t]
\caption{Imaging parameters used for producing the high-resolution clean images (High Res Clean) and the dirty image cubes for the power spectrum (Dirty Cube). The values of parameters not applicable for making dirty images are specified as "$\cdots$".}
\label{tab:imaging}
\centering
\begin{tabular}{@{}lll@{}}
\toprule
Parameter      & High Res Clean            & Dirty Cube       \\ \midrule
Weighting      & Briggs (Robust = --0.1)    & Natural          \\
Pixel size     & 3\arcmin                        & 6\arcmin               \\
Image size     & 600 pixels                & 400 pixels       \\
$u\varv$ range       & >20$\lambda$              & 5--100$\lambda$  \\
Channels       & 12 (joined)               & 177 (not joined) \\
Algorithm      & Multiscale (Hogbom)       & $\cdots$              \\
Auto mask\tablefootmark{a}      & 3                         & $\cdots$              \\
Auto threshold\tablefootmark{b} & 1                         & $\cdots$              \\
Mgain\tablefootmark{c}          & 0.6                       & $\cdots$              \\
Spectral pol\tablefootmark{d}   & 2 terms                 & $\cdots$              \\ \bottomrule
\end{tabular}
\tablefoottext{a}{Threshold at which a mask is constructed from the obtained components.}
\tablefoottext{b}{Cleaning threshold based on the local rms.}
\tablefoottext{c}{Cleaning gain for major iterations.}
\tablefoottext{d}{Number of terms in the polynomial fit over frequency to each clean component.}
\end{table}

\begin{figure*}
        \includegraphics[width=\hsize]{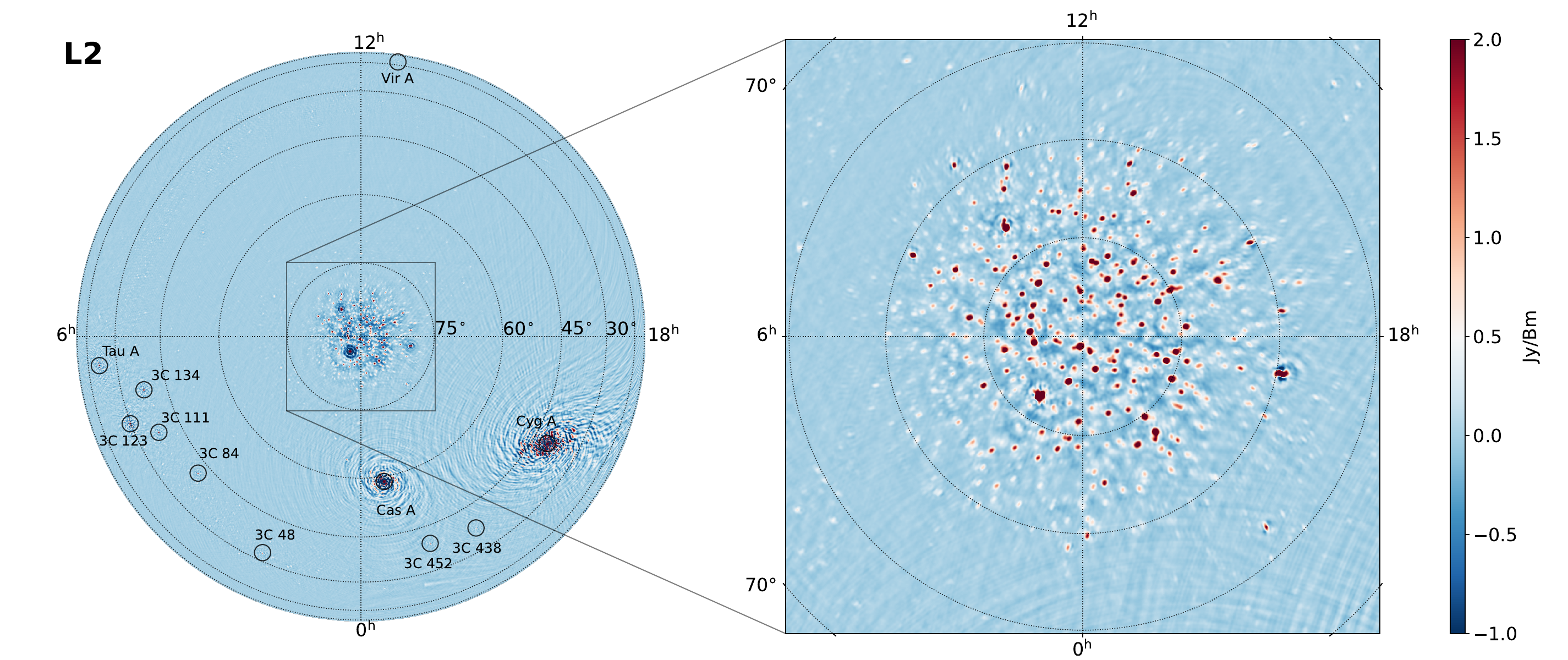}
	
        \includegraphics[width=\hsize]{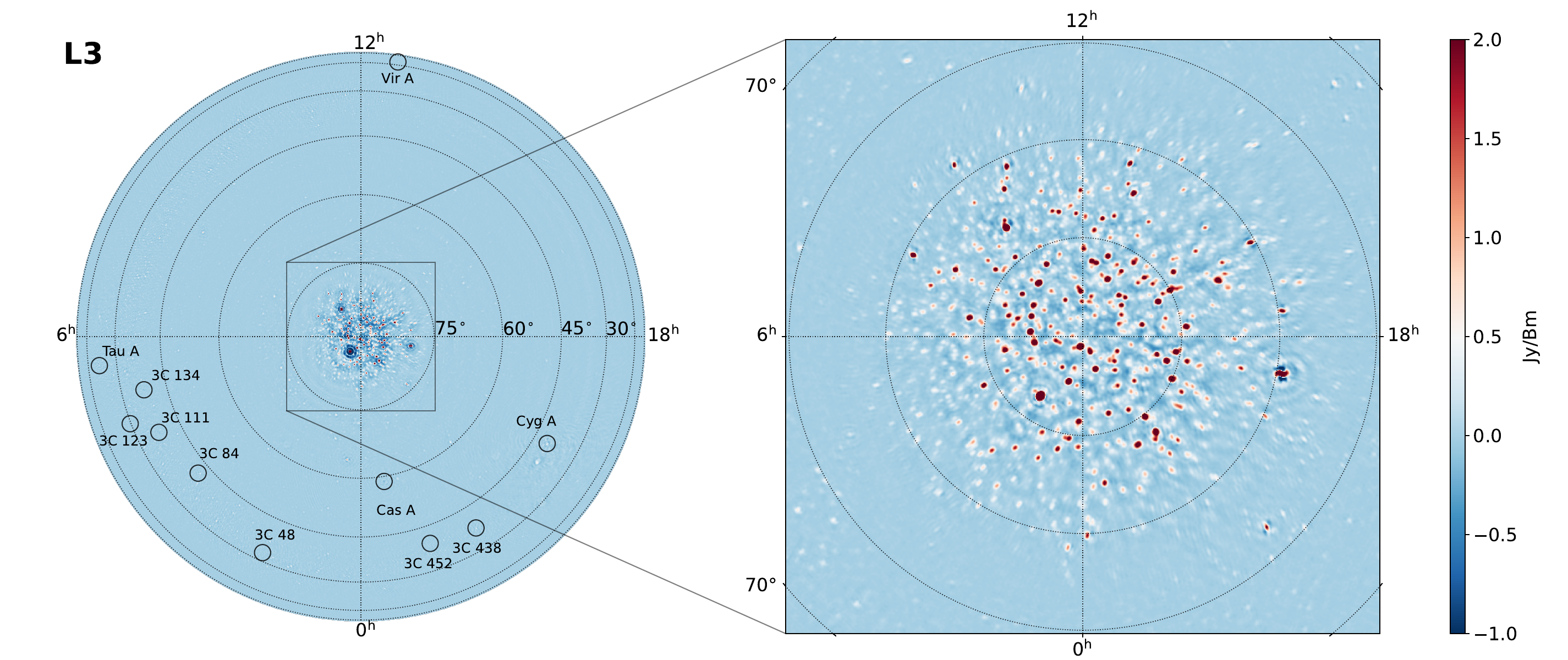}
	
        \includegraphics[width=\hsize]{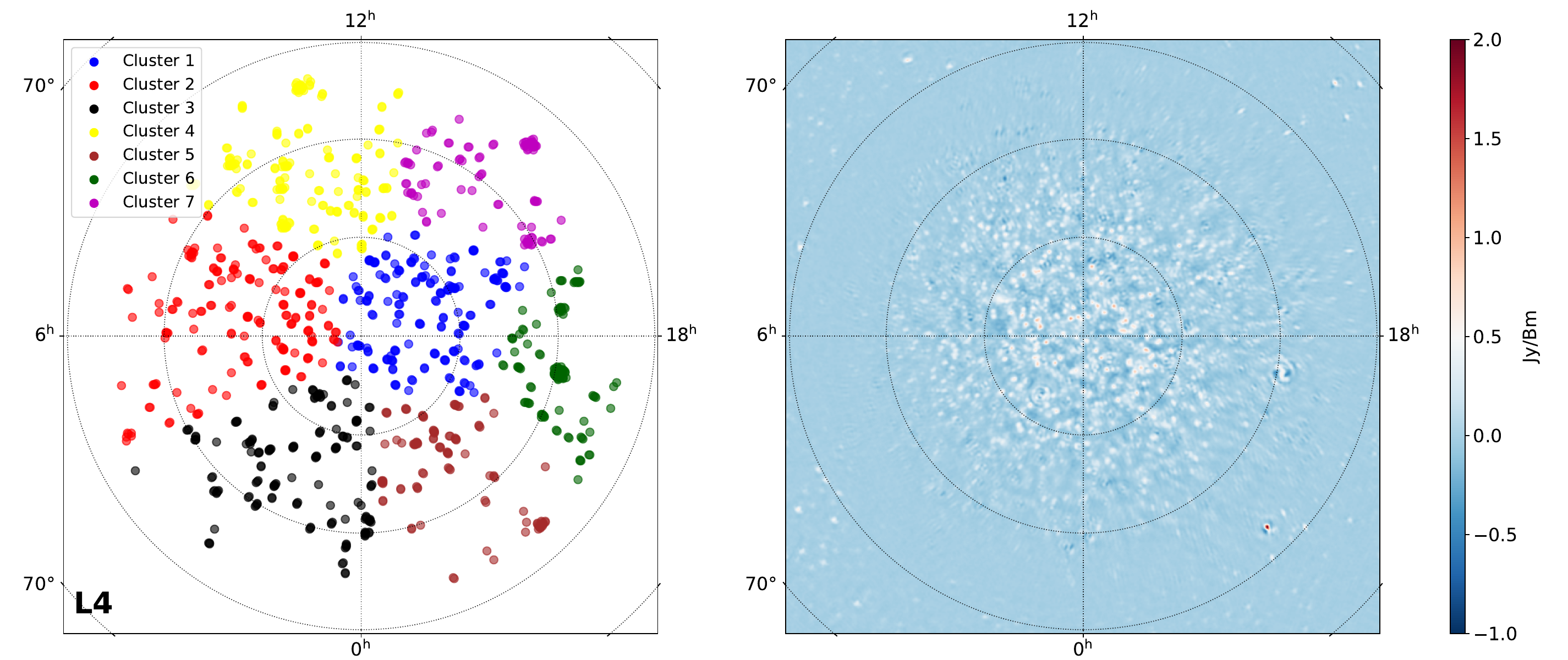}
    \caption{Images of the L2, L3, and L4 data. Top: Wide-field dirty image (left) and narrow-field cleaned image (right) of the L2 data. Middle: Same but for the L3 data. Bottom left: The seven clusters that the sky model is divided into for the NCP subtraction step. Bottom right: Narrow-field cleaned image of the L4 data.}
    \label{fig:clean_images}
\end{figure*}

\begin{figure*}
        \includegraphics[width=\hsize]{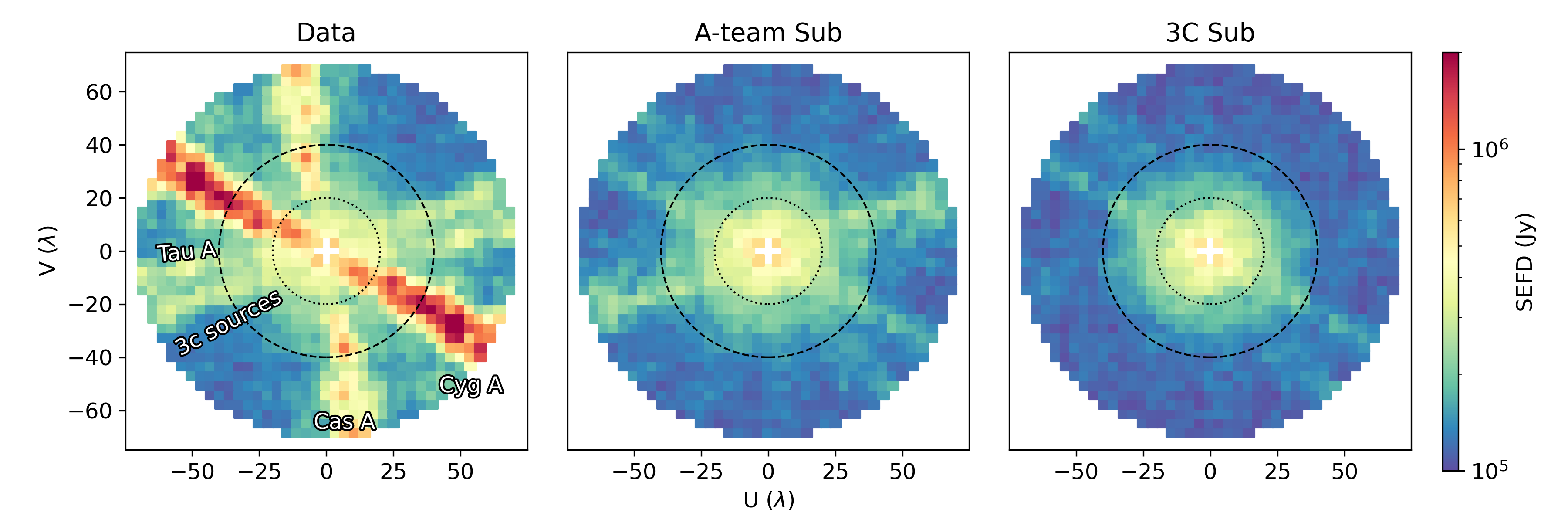}
    \caption{Data standard deviation in SEFD units calculated from the channel-differenced noise at each $u\varv$ cell of the Stokes-I data cube. The three panels correspond to the data after preprocessing (left), after A-team subtraction (middle), and after 3C subtraction (right). The dotted and dashed circles correspond to the 20$\lambda$ and 40$\lambda$ limits, respectively, between which the power spectrum is finally computed.}
    \label{fig:uvfeatures}
\end{figure*}

To assess the level of residuals at each stage of calibration and subtraction, the visibilities at each stage were imaged using {\tt WSClean} and the MFS images of the data at different stages are presented in Fig. \ref{fig:clean_images}. The left panels of Fig. \ref{fig:clean_images} show dirty wide-field images, which were made with the same parameters reported in the "High Res Clean" column of Table \ref{tab:imaging}, except with an image size of 2400 pixels and without the last five parameters, which are needed only for cleaning. The wide-field image in the top-left panel clearly shows the level of contamination from the A-team sources (particularly Cas A and Cyg A), which are well subtracted in the A-team subtraction step (mid-left panel). The narrow-field cleaned images (right panels of Fig. \ref{fig:clean_images}) show how the PSF sidelobes of the A-team running through the NCP field, are also well subtracted. To assess the impact of the PSF sidelobes of wide-field sources on the $u \varv$ plane, the standard deviation of the data in system equivalent flux density (SEFD) units was calculated from the frequency-channel-differenced noise for each $u \varv$ cell in the gridded data cube (the gridding procedure is described in Sects. \ref{sec:image_cubes} and \ref{sec:viscube}). This was repeated at different stages of source subtraction and the results are presented in Fig. \ref{fig:uvfeatures}. It should be noted that only the central 16 degrees of the field was used in constructing these data cubes as well as the final power spectra. This figure illustrates the impact of the PSF sidelobes of the A-team sources, which appear as straight lines in the spatial frequency domain\footnote{These lines are perpendicular to the ripples of the PSF sidelobes of sources located far away from the phase center.} and are well subtracted using this method. However, the subtraction of Cyg A is not perfect and the PSF sidelobes of the Cyg A residuals can still be seen after the A-team subtraction step (middle and right panels of Fig. \ref{fig:uvfeatures}).

\subsubsection{Subtraction of 3C sources}
In addition to the A-team, there are several other compact sources in the northern sky that have fluxes of hundreds of janskys at the low frequencies of interest in our observations. In general, most of these sources will be highly attenuated by the primary beam and hence their PSF sidelobes will not be bright enough to contaminate the NCP field. However, in our data, we find that the PSF sidelobes of seven sources\footnote{3C\,123, 3C\,134, 3C\,111, 3C\,84, 3C\,48, 3C\,452, and 3C\,438} do have a strong effect on the NCP field. This is because all these sources pass through the grating lobes of the MA beam, and the primary attenuation that their fluxes get is due to the dipole beam and not due to the array factor of the MA, which is very high at the grating lobes. From Fig. \ref{fig:uvfeatures}, it is clear that the PSF sidelobes of the 3C sources have a strong signature on the $u\varv$ plane. To subtract these sources, we performed a DD calibration in eight directions: the seven 3C sources and the NCP. The time solution interval of 4 min was selected to be small enough to allow the gain solutions to model the grating lobes and large enough such that there is sufficient signal-to-noise ratio (S/N) to obtain reliable calibration solutions. The sky model used for each of these sources is a single delta function at the location of the source, obtained from catalogs. The sources were included in the sky model for all segments where they are above 10 degrees altitude. This results in 3C\,438 being excluded for the last seven segments, 3C\,452 for the last six segments, 3C\,48 for the last two segments, and 3C\,84 for the last segment. The three other sources were included in the calibration for all segments. The third iteration NCP sky model made from the A-team subtracted and DI-corrected data was used as the model in the direction of the NCP field. Once the calibration solutions were obtained, the visibilities for the 3C sources were predicted, the calculated gains in the respective direction were applied and the corrupted predicted visibilities were subtracted from the data. This largely removes these seven sources and their PSF sidelobes from the data. The impact of this step on the image and the  $u\varv$ plane are shown in Figs. \ref{fig:clean_images} and \ref{fig:uvfeatures}, respectively. The impact of the PSF sidelobes is particularly prominent in Fig. \ref{fig:uvfeatures} and the sidelobes are seen to be well subtracted using this method (right panel). The bright source subtracted and DI-corrected data will hereafter be referred to as the L3 data.

\begin{table}[t]
\caption{NCP sky model used at different analysis stages.}
\label{tab:sm}
\begin{tabular}{@{}llccc@{}}
\toprule
Sky Model & Direction & Components & Total Flux & Max Flux \\ 
    &   &   & (Jy) & (Jy)\\
\midrule
Initial\tablefootmark{*}   & NCP       & 2787\tablefootmark{*}       & 6379\tablefootmark{*}     & 83\tablefootmark{*}    \\
2nd Iteration & NCP       & 1608       & 726     & 40    \\
3rd Iteration & NCP       & 1622       & 733     & 39    \\ \midrule
Clustered & Cluster 1 & 191        & 221     & 5     \\
          & Cluster 2 & 182        & 144     & 7     \\
          & Cluster 3 & 163        & 127     & 39    \\
          & Cluster 4 & 186        & 88      & 12    \\
          & Cluster 5 & 107        & 58      & 4     \\
          & Cluster 6 & 124        & 36      & 8     \\
          & Cluster 7 & 95         & 33      & 3     \\ \bottomrule
\end{tabular}
\tablefoottext{*}{The "Initial" model is an intrinsic sky model, while all other models are apparent sky models.}
\end{table}

\subsection{NCP subtraction}\label{sec:ncp_sub}
The sources in the NCP field can now be subtracted from the L3 data. Here it is important to account for the contaminating effects of ionospheric phase shifts and a non-axisymmetric central lobe of the primary beam of NenuFAR, both of which are expected to introduce DD effects on the data within the field of view (FOV). Currently, it is not possible to solve for small timescale ionospheric effects because the solution time interval needed to have a high enough S/N to yield reliable calibration solutions is much longer than the timescale on which these effects occur on short baselines (a few seconds to 1 min, following \citet{vedantham2015scintillation}). Ideally one would want to obtain separate gain matrices for all the different components in the NCP sky model. However, this is infeasible because the flux for each component is not sufficient to yield reliable calibration solutions and it would also be extremely computationally expensive. Additionally, solving for a large number of directions would also increase the degrees of freedom, thus increasing the risk of absorption of the noise and the 21 cm signal. Therefore, we used an alternative approach, employed by the LOFAR EoR KSP, in which the NCP sky model is divided into several clusters and a separate calibration solution is obtained for each cluster under the assumption that DD effects do not vary strongly over each cluster \citep{kazemi2013clustered}. For this purpose, we used the third iteration sky model made from the data after DI correction (Table \ref{tab:sm}). The sources outside the primary beam beyond 13.4 degrees from the phase center were first removed from the sky model since they have a low flux density. The remaining sources were divided into seven clusters using a $k$-means-based clustering algorithm. The number seven was chosen to maximize the number of clusters and retain a sufficiently large S/N for the faintest cluster. The flux in the faintest cluster is 33 Jy, which is at a 6$\sigma$ level when compared to the expected thermal noise in the visibilities for the chosen 8 min and 183.1~kHz solution interval. The details of the clustered sky model are presented in the bottom half of Table \ref{tab:sm} and the clusters are shown in the bottom-left panel of Figure \ref{fig:clean_images}. The L3 data were then calibrated against this clustered sky model, with separate solutions calculated for the direction of each cluster. Once the calibration solutions were obtained, the visibilities corresponding to each cluster were predicted, then corrupted by the calculated gains and finally subtracted from the data. We performed multiple runs of the DD calibration in conjunction with signal injection simulations using a fiducial simulated 21 cm signal (described in Sect. \ref{sec:robustness_ddcal}) to converge to the final calibration settings, which gives the best compromise between optimal results in calibration solutions and images and minimum signal suppression. The data after NCP subtraction will hereafter be referred to as the L4 data.

In the bottom-right panel of Fig. \ref{fig:clean_images} we can see that most sources are subtracted well until we reach the confusion noise level (of 0.9~Jy/Beam\footnote{This is estimated using Eq. 6 of \citet{van2013lofar}.}) beyond which we do not have sky model components and hence they are not subtracted. The rms of the fluxes in the image pixels in the central 5 degrees radius is 0.15~Jy/Beam, with a maximum and minimum flux of 1.05~Jy/Beam and $-0.45$~Jy/Beam, respectively. The remaining poorly subtracted sources at the edge of the clusters are likely due to the fact that the DD effects are not accounted for on scales smaller than the cluster sizes. As a result, the average gain solution for each cluster does not correctly represent the gains in the direction of these sources and is a complex flux-weighted average over the cluster.

\section{Post-calibration RFI flagging}\label{sec:postflag}
\begin{figure}
        \includegraphics[width=\hsize]{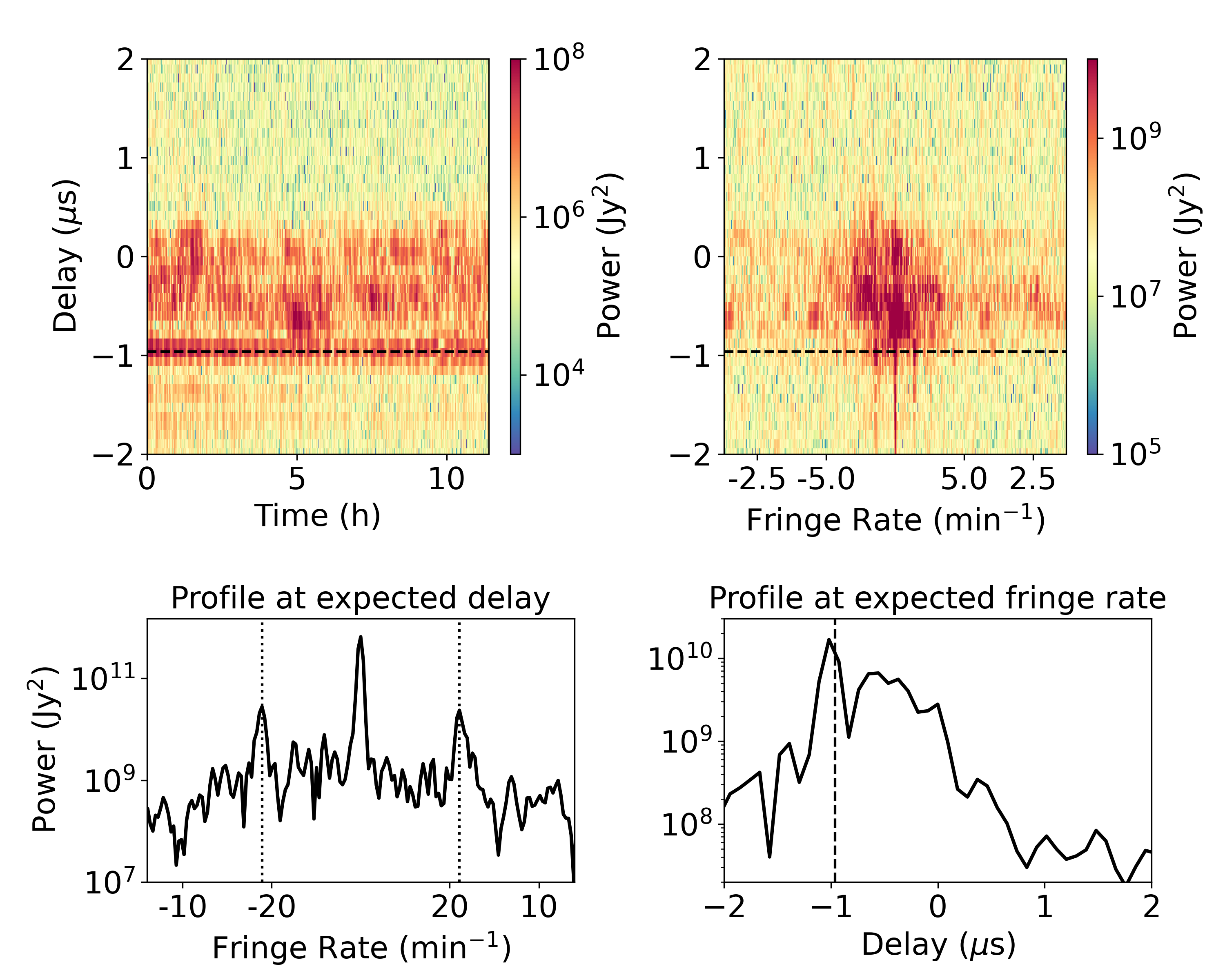}
    \caption{Delay and fringe rate power spectra of an example baseline where a strong RFI feature is seen. Delay vs. time plot (top-left) and delay vs. fringe rate plot (top-right) for the Stokes-I data after NCP subtraction. The dashed horizontal lines indicate the expected delay corresponding to the local RFI source at the building. Bottom left: Power as a function of fringe rate at the expected delay for the local RFI source. The vertical dotted lines correspond to a fringe rate of 18 $\textrm{min}^{-1}$. Bottom right: Power as a function of delay at the expected fringe rate of 18 $\textrm{min}^{-1}$. The vertical dashed line is the expected delay.}  
    \label{fig:delay_fringe}
\end{figure}

The data after the different stages of calibration and point source subtraction still have significant residual power at high delay (Fourier dual of frequency) modes. Strong features are seen in the delay and fringe rate (Fourier dual of time) power spectra of multiple baselines (Fig. \ref{fig:delay_fringe}), which are likely due to near-field RFI.

\begin{figure*}
        \includegraphics[width=\hsize]{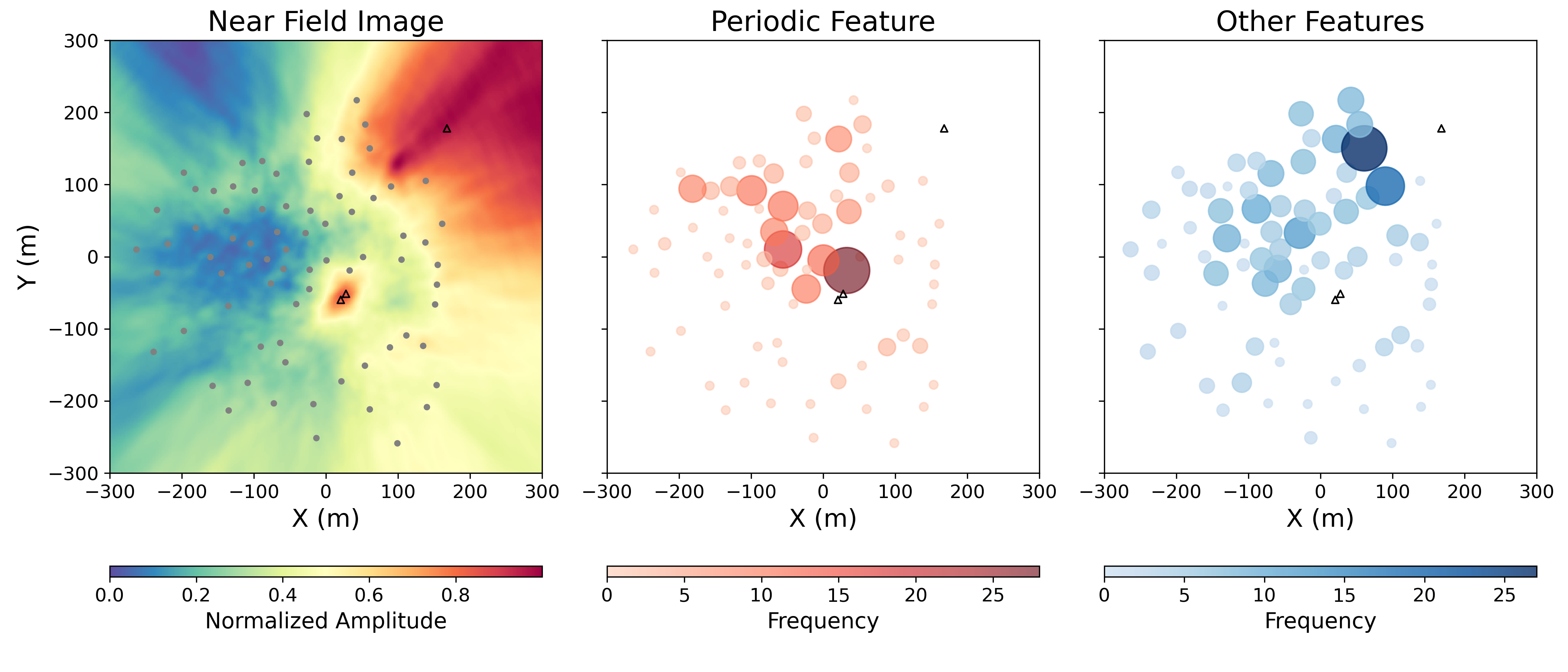}
    \caption{Local RFI sources at the NenuFAR site and their impact on the MAs. Left: Normalized near-field image of the data. The locations of the buildings housing the electronic containers are indicated with black triangles, and the MAs are indicated with gray circles. Center: Histogram of the MAs contributing to the baselines that exhibit a periodic amplitude fluctuation in time. Right: Histogram of the MAs contributing to all other baselines, which have been flagged based on unusual features seen in the delay power spectra. The size and color of the circles in the middle and right panels indicate the number of baselines involving that MA that have been flagged.}
    \label{fig:nfi}
\end{figure*}
To identify local sources of RFI, we constructed a near-field image \citep{paciga2011gmrt} from the L4 data. This image was created by coherently summing all visibilities after their amplitudes had been set to unity and assuming that phase differences are only due to sources on the ground. Thus, the amplitude in the image does not necessarily correspond to the strength of the local RFI source. It only allows us to pinpoint its location. The near field image reveals that the buildings within the NenuFAR core, housing the electronic containers, are producing significant RFI (left panel of Fig. \ref{fig:nfi}). We performed a comprehensive study of the impact of such near-field sources of RFI on the power spectrum. The details of this study will be published in a separate paper. To mitigate this near-field RFI, we adopted a simple approach of examining the delay power spectra of individual baselines and selecting baselines that are most severely affected by the RFI source. The RFI source at the two buildings within the array is seen to have a periodic fluctuation in intensity with a periodicity of 18 min, leading to a well-defined signature in the delay-fringe rate space (Fig. \ref{fig:delay_fringe}). The origin of this periodicity is still unknown. Any baseline showing a similar feature in the delay-fringe rate power spectra can be easily identified and flagged. This can potentially allow us to filter it out from the data instead of manually flagging entire baselines, but we defer this to future analysis. We examined the Stokes-I delay power spectra for all baselines manually and identified those that show a strong periodic feature. For each such baseline, a feature is always also present in the Stokes-V delay spectra, where it is usually more prominent because the power from sky sources is negligible. The histogram of the MAs contributing to the flagged baselines reveals that baselines involving MAs close to the building are more strongly affected by the RFI (middle panel of Fig. \ref{fig:nfi}). In addition to this periodic RFI signature, the delay spectra of many baselines also show other unusually strong features beyond the horizon delays. The histogram of the MAs contributing to these baselines shows a peak near the northeast of the array (right panel of Fig. \ref{fig:nfi}) in a region where we see a strong RFI source in the near-field image. These baselines were flagged as well. In this process, we flagged 7.5$\%$ of the data. After this additional flagging step, the data are deemed sufficiently clean, calibrated and sky model subtracted, and can be used for residual foreground subtraction.

\section{Power spectra estimation}\label{sec:ps}
The estimation of power spectra was done using the power spectrum pipeline {\tt pspipe},\footnote{\url{https://gitlab.com/flomertens/pspipe}} which is also used to generate the power spectrum for the LOFAR EoR KSP.

\subsection{Image cubes}\label{sec:image_cubes}
The data need to be gridded in a $u\varv\nu$ grid in order to construct the power spectra. For this purpose, {\tt WSClean} was used to construct dirty image cubes from the visibilities. {\tt WSClean} uses a $w$-stacking algorithm while making the image, which accounts for wide-field effects due to the $w$-term. The imaging parameters are specified in the column named "Dirty Cube" in Table \ref{tab:imaging}. Separate image cubes were made for alternating odd and even time samples and these were used at a later stage in the estimation of the noise level in the data. A Hann filter with an FOV of 16 degrees was applied to the dirty image cubes in the image plane in order to suppress primary-beam effects on sources far away from the phase center as well as aliasing artifacts. These dirty image cubes were subsequently used to obtain the power spectra.

\subsection{Conversion to visibility cubes}\label{sec:viscube}
The gridded image cubes ($I^{\text{D}}$) produced as described above were first converted from units of Jy/PSF to units of Kelvin using the relation \citep{offringa2019impact,mertens2020improved}\begin{equation}
T(\textit{l,m},\nu) = \dfrac{10^{-26}c^{2}}{2k_{\text{B}}\nu^{2}\delta_{l}\delta_{m}}\mathcal{F}^{-1}_{u,\varv}[\mathcal{F}_{\textit{l,m}}[I^{\text{D}}]\oslash\mathcal{F}_{\textit{l,m}}[I^{\text{PSF}}]],
\end{equation}
where $k_{\text{B}}$ is the Boltzmann constant, $\delta_{l}, \delta_{m}$ are the pixel sizes in the $l$ and $m$ directions (in units of radians), respectively, $\mathcal{F}_{\textit{l,m}}$ denotes a Fourier transform and $\mathcal{F}^{-1}_{u,\varv}$ is its inverse, $I^{\text{PSF}}$ is the point spread function and $\oslash$ is the element-wise division operator. These image cubes in the $(\textit{l,m},\nu)$ space were then Fourier transformed in the spatial direction to get the data cube in the $u\varv\nu$-space: $\tilde{T}(u,\varv,\nu)$. A $u\varv$ range of 15 to 50 $\lambda$ was chosen for further analysis, and the remaining data were nulled for all frequencies. This avoids baselines shorter than 15$\lambda,$ which could have strong contaminating effects from mutual coupling, and those longer than 50 $\lambda,$ which have a higher thermal noise contribution. We note that though we constructed the final spherical power spectra in the $20-40\lambda$ baseline range, here we tried to retain as many baselines as possible since it allows better foreground modeling and removal through ML-GPR. A final outlier rejection was done on the gridded visibility cubes to flag potentially remaining low-level RFI using a simple threshold-clipping method. The $u\varv$ cells were flagged based on outliers in the $u\varv$ weights, Stokes-V variance, and channel-differenced Stokes-I variance. About 15$\%$ of the $u\varv$ cells were flagged in this procedure.

\begin{figure*}
        \includegraphics[width=\hsize]{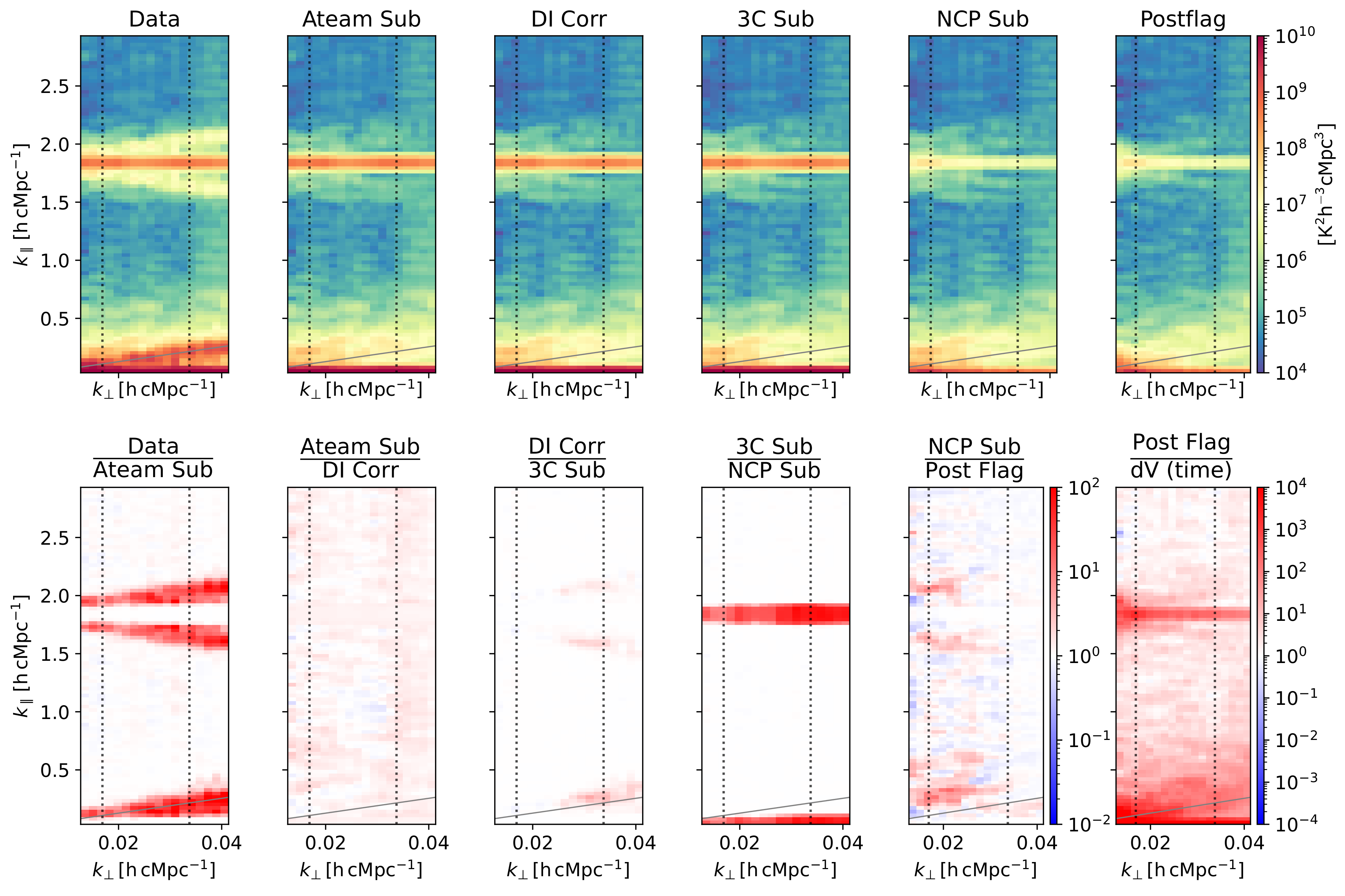}
    \caption{Cylindrical power spectra at different stages of calibration and source subtraction. Top row: Cylindrical power spectra after preprocessing ("Data"), after A-team subtraction ("Ateam Sub"), after DI correction ("DI Corr"), after 3C subtraction ("3C Sub"), after NCP Subtraction ("NCP Sub"), and after post-calibration RFI flagging ("Postflag"). Bottom row: Ratio of successive power spectra in the top row. This shows how much power has been subtracted and from which part of the $k$ space at the different calibration stages. The rightmost plot in the bottom row is the ratio of the power spectrum of the data after post-calibration RFI flagging to the noise power spectrum obtained from time-differenced Stokes-V data. Note that the color bars for all plots in the same row are the same, except for the plot on the extreme right in the bottom row.}
    \label{fig:ps2d}
\end{figure*}

\subsection{Cylindrically and spherically averaged power spectra}\label{sec:2d1dps}
The data cube $\tilde{T}(u,\varv,\nu)$ at any stage of calibration and foreground subtraction can now be Fourier transformed along the frequency axis (after applying a Blackman-Harris filter to suppress aliasing) to obtain $\tilde{T}(u,\varv,\eta),$ where $\eta$ is the Fourier dual of frequency, commonly referred to as the delay. The corresponding power spectrum as a function of the wave numbers $k_{l},k_{m}, k_{\parallel}$ is given by
\begin{equation}\label{eq:temp_to_power}
P(k_{l},k_{m},k_{\parallel}) = \dfrac{\Omega_{lm}B_{\text{bw}} D^{2}_{M}(z)\Delta D}{\langle A^{2}_{\text{pb}}(\textit{l,m})A^{2}_{w}(\textit{l,m})\rangle \langle B^{2}_{w}(\nu)\rangle}|\tilde{T}(u,\varv,\eta)|^{2}.
\end{equation}
The wave modes are related to $u,\varv,\eta$ as \citep{morales2004toward,mcquinn2006cosmological}\begin{equation}
k_{l} = \dfrac{2\pi u}{\text{D}_{M}(z)}, \; k_{m} = \dfrac{2\pi \varv} {\text{D}_{M}(z)}, \; k_{\parallel} = \dfrac{2\pi H_{0}\nu_{21}E(z)}{c(1+z)^{2}}\eta,
\end{equation}
where $\nu_{21}$ = 1420~MHz, $H_{0}$ is the Hubble constant, $E(z)$ is the dimensionless Hubble parameter with $E(z) = H(z)/H_{0}$ where $H(z)$ is the Hubble parameter.
$\Omega_{lm}$ denotes the angular extent of the image cube and $B_{\text{bw}}$ denotes the frequency bandwidth of the data cube. $\text{D}_{M}(z)$ and $\Delta \text{D}$ are conversion factors to go to comoving distance units from angular and frequency units, respectively. The denominator in Eq. \ref{eq:temp_to_power} accounts for the limitation of the angular extent due to the primary beam $A_{pb}(\textit{l,m})$ and the spatial tapering function $A_{w}(\textit{l,m})$ and also the limitation in the frequency extent due to the frequency filter $B_{w}(\nu)$ used before the Fourier transform along the frequency direction. $\langle \cdots \rangle$ denotes an average over the respective domains. The power values $P(k_{l},k_{m},k_{\parallel})$ are next averaged in cylindrical and spherical shells to yield the cylindrical (two-dimensional) power spectrum $P(k_{\perp},k_{\parallel})$ and the spherical power spectrum $P(k)$, respectively, where $k_{\perp}^2 = k_{l}^{2}+k_{m}^{2}$ and $k^2=k_{\perp}^{2}+k_{\parallel}^{2}$.

We computed the Stokes-I cylindrical power spectra at all the stages of calibration as a diagnostic to compare the power levels after each step. The cylindrical power spectra for each calibration step and the ratio of the power spectra in successive steps are presented in Fig. \ref{fig:ps2d}. The gray lines correspond to the horizon delays and the foreground wedge is clearly visible to have several orders of magnitude higher power than the "EoR window." The bright horizontal feature at $k_{\parallel} \approx 1.84\ h\, \text{cMpc}^{-1}$ is the result of the flagging of sets of 2 channels at the ends of each sub-band during preprocessing, in order to avoid the edge effect of the polyphase filter used in forming the sub-bands. This feature at such a high $k$ mode does not affect our power spectrum analysis significantly, since it is focused on $k$ modes typically much smaller than this. It is evident that a significant amount of power near the wedge's horizon is removed during the A-team and 3C subtraction steps, while the NCP subtraction removes power at the lowest $k_{\parallel}$ modes corresponding to the low delay values of the NCP field at the phase center. Post-calibration flagging decreases the power by a factor of more than 10 in the high $k_{\parallel}$-modes beyond the horizon line ($0.2-0.6$ $h\, \text{cMpc}^{-1}$). In some $k$-modes this factor is less than unity because the thermal noise level is higher after flagging due to the lower volume of remaining data. It should be noted that the data prior to the DI correction stage is not absolutely calibrated. The noise power spectrum is estimated by taking the difference of the Stokes-V dirty image cubes of the even and odd time samples and then forming the power spectra as usual, accounting for the extra factor of 2 increase in variance in the process. The panel on the bottom right in Fig. \ref{fig:ps2d} shows the power spectrum of the data after post-calibration RFI flagging divided by the noise power. We see that within the wedge, there is still more than three orders of magnitude of power beyond the thermal noise limit, likely due to Galactic diffuse emission and confusion noise due to extragalactic sources. The power far beyond the wedge could be due to a variety of factors such as residual RFI and polarization leakage. Well away from the wedge, at $k_{\parallel}>1\ h\, \text{cMpc}^{-1}$, the residual power approaches the thermal noise limit.

\section{Residual foreground removal}\label{sec:mlgpr}
The data, after the different steps of calibration and compact source subtraction, are still dominated by residual foregrounds, such as the diffuse Galactic emission and extragalactic point sources, which have a flux density at or below the confusion noise limit. However, the fact that foregrounds have a larger frequency coherence scale than thermal noise or the 21 cm signal can be utilized to model and subtract the foregrounds from the data. One approach for subtracting foregrounds from the data is through GPR \citep{mertens2018statistical}. 

\subsection{Gaussian process regression}
In GPR, the data are modeled as a sum of Gaussian processes describing the foreground, thermal noise, and the 21 cm signal components. Each Gaussian process is characterized by a certain frequency covariance function and zero mean. The covariance function is parametrized by a set of adjustable hyperparameters that control properties such as the variance, coherence scale, and the shape of the covariance function. Using a Bayesian approach, the maximum a posteriori values of the hyperparameters are derived from their posterior probability distribution conditioned on the observed data. The expectation value of the foreground component at each data point is subtracted from the data to yield the residual foreground subtracted data.

To limit the computational requirements, we applied GPR to the gridded visibility data cube $\tilde{T}(u,\varv,\nu)$ before power spectrum generation. Performing GPR along the frequency direction in the $u\varv$ space allows us to easily take into account the baseline dependence of the frequency coherence scale of the different components, such as the foreground wedge and the thermal noise. Following the same approach as taken by \citet{mertens2018statistical}, the data \textbf{d} is modeled as a sum of different components, namely the foregrounds $\textbf{f}_{\mathrm{fg}}$, the 21 cm signal $\textbf{f}_{21}$ and the noise \textbf{n} as functions of frequency $\nu$,
\begin{equation}\label{eq:gp_data}
\textbf{d} = \textbf{f}_{\mathrm{fg}}(\nu) + \textbf{f}_{21}(\nu) + \textbf{n}(\nu).
\end{equation}
The different GP components should in principle be possible to separate by virtue of their having different spectral behavior. This spectral behavior is specified by the covariance of the components between frequency channels, with the total covariance matrix of the data being a sum of the individual GP covariances:

\begin{equation}\label{eq:gp_covariances}
   \textbf{K}(\nu_{p},\nu_{q}) = \textbf{K}_{\mathrm{fg}}(\nu_{p},\nu_{q}) + \textbf{K}_{21}(\nu_{p},\nu_{q}) + \textbf{K}_{n}(\nu_{p},\nu_{q}).
\end{equation}
Here \textbf{K} is the total covariance matrix of the data whose entries are a function of two frequencies $\nu_{p}$ and $\nu_{q}$, and is given by the sum of the foregrounds covariance matrix $\textbf{K}_{\mathrm{fg}}$, the 21 cm covariance matrix $\textbf{K}_{21}$ and the noise covariance matrix $\textbf{K}_{n}$. 

In an earlier approach to implementing GPR, which was employed by \citet{gehlot2019first}, \citet{mertens2020improved}, and \citet{gehlot2020aartfaac}, both the foreground and 21 cm covariances have a specific functional form along with hyperparameters that guide their respective variance and frequency coherence scale. However, one concern with this method is that there is a risk of signal loss if the 21 cm covariance function is not a good match to the frequency covariance of the 21 cm signal in actual data \citep{kern2021gaussian}. This becomes particularly important when the objective is to use the 21 cm power spectrum upper limits to rule out astrophysical models that have a 21 cm power spectrum that is not well described by the covariance function adopted for the 21 cm signal in GPR. Addressing these concerns, a novel approach to GPR-based foreground subtraction called ML-GPR has been developed by \citet{mertens2023retrieving}, which employs ML methods to build a covariance model of the 21 cm signal directly from simulations. The main steps in the ML-GPR approach are summarized in the following section.

\begin{figure}
    \includegraphics{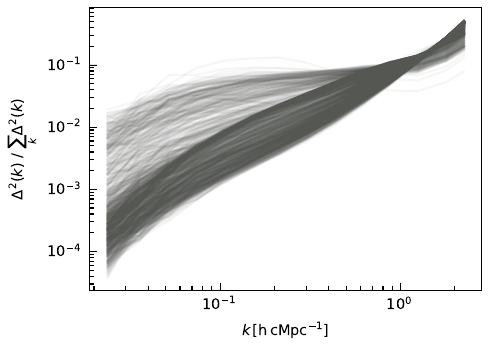}
    \caption{Normalized spherically averaged power spectra of the 1000 {\tt 21cmFAST} simulations at $z$=20 used as the training set for the 21 cm VAE kernel.}
    \label{fig:all_ps3d_simu}
\end{figure}

\begin{table*}[t]
\caption{Components of the covariance model used in ML-GPR along with the priors and converged values of the parameters.}
\label{tab:mlgpr}
\centering
\begin{tabular}{@{}llclcc@{}}
\toprule
Component               & Covariance            & Parameter          & Description            & Prior Bounds  & Estimated Value   \\ \midrule\midrule
Intrinsic foregrounds   & Radial Basis Function & $\sigma^{2}_{\text{int}}$ & Variance               & {[}--1,1{]}    & $-0.1 \pm 0.02$   \\
                        &                       & $l_{\text{int}}$          & Lengthscale            & {[}20,40{]}   & $32.9 \pm 1.4$    \\ \midrule
Mode-mixing foregrounds & Radial Basis Function & $\sigma^{2}_{\text{mix}}$ & Variance               & {[}--2,0{]}    & $-1.32 \pm 0.01$  \\
                        &                       & $l_{\text{mix}}$          & Lengthscale            & {[}0.1,0.5{]} & $0.275 \pm 0.001$ \\ \midrule
21 cm signal            & Trained ML Kernel     & $x_1$              & Latent space dimension & {[}--4,4{]}    & -                 \\
                        &                       & $x_2$              & Latent space dimension & {[}--4,4{]}    & -                 \\
                        &                       & $\sigma^{2}_{21}$  & Variance               & {[}--7,--1{]}   & $< -4.97$         \\ \midrule
Excess power            & Exponential Function  & $\sigma^{2}_{\text{ex}}$  & Variance               & {[}--3,--1{]}   & $-1.88 \pm 0.02$  \\
                        &                       & $l_{\text{ex}}$           & Lengthscale            & {[}0.2,2{]}   & $0.56 \pm 0.03$   \\ \bottomrule
\end{tabular}
\tablefoot{All $\sigma^{2}$ values are in logarithmic scale and are expressed as a fraction of the variance of the data input into GPR. $\sigma^{2}_{21}$ reaches the lower bound and hence only the upper limit is shown. The priors are all uniform priors. All \textit{l} values are in units of MHz.}
\end{table*}

\subsection{Training}
In ML-GPR, a variational auto-encoder (VAE) kernel is trained to build a low dimensional representation of the 21 cm signal covariance from a large number of simulations of the 21 cm signal. We used the {\tt 21cmFAST}~\citep{mesinger201121cmfast, Murray2020} code to produce simulations of the 21 cm signal at our redshift of interest ($z=20$), with a comoving box size of 600 Mpc and a resolution of 2 Mpc per pixel. We followed the parameterization introduced by~\citet{park2019inferring} with the following range of astrophysical parameters:
\begin{itemize}[leftmargin=40pt]
    \setlength{\itemsep}{0.25em}
    \item[$f_{*,10}$] the normalization of the fraction of Galactic gas in stars at high $z$, evaluated for halos of mass $10^{10}M_{\odot}$: $\mathrm{log_{10}(f_{*,10}) = [-3, 0]}$.
    
    \item[$\alpha_\ast$] the power-law scaling of $f_\ast$ with halo mass: $\alpha_\ast = [-0.5, 1]$.
    
    \item[$f_{\mathrm{esc},10}$] the normalization of the ionizing UV escape fraction of high $z$ galaxies, evaluated for halos of mass $10^{10}M_{\odot}$: $\mathrm{log}_{10}(f_{\mathrm{esc},10}) = [-3,0]$.

    \item[$\alpha_{\mathrm{esc}}$] the power-law scaling of $f_{\mathrm{esc}}$ with halo mass: $\alpha_{\mathrm{esc}} = [-0.5, 1]$.

    \item[$t_*$] the star formation timescale taken as a fraction of the Hubble time: $t_{*} = [0, 1]$.

    \item[$M_{\mathrm{turn}}$] the turnover halo mass below which the abundance of active star forming galaxies is exponentially suppressed: $\mathrm{log_{10}} (M_{\mathrm{turn}}/M_{\odot}) = [8, 10]$.

    \item[$E_0$] the minimum X-ray photon energy capable of escaping the galaxy, in keV: $E_0 = [0.1, 1.5]$.

    \item[$L_{\mathrm{X}}\mathrm{/SFR}$] the normalization of the soft-band X-ray luminosity per unit star formation computed over the band $E_0 - 2$ keV.: $\mathrm{log_{10}}(L_{\mathrm{X}}\mathrm{/SFR}) = [38, 42]$.
\end{itemize}

Generating meaningful simulations of cosmic dawn made it necessary to perform IGM spin temperature fluctuations. Inhomogeneous recombination was also turned on in the simulations. Latin hypercube sampling was used to sample 1000 sets of parameters in this eight-dimensional space and {\tt 21cmFAST} was used to perform 1000 simulations and obtain the corresponding brightness temperature cubes. We note that this relatively sparsely sampled parameter space is sufficient since the VAE, being a generative model, is able to interpolate between the training samples. This has been shown by \citet{mertens2023retrieving}. Additionally, since we used a VAE instead of an auto-encoder (AE), the added regularization allowed us to avoid overfitting the sparse sample \citep{kingma2013auto}. Next, the spherically averaged power spectra for all 21 cm brightness temperature cubes were computed and normalized in the $k$ range $0.03-2.0\ h\,\text{cMpc}^{-1}$ to have a variance of unity. This was done because we aim to use the VAE to learn only the shape of the 21 cm power spectrum and keep its variance as a separate free parameter, thus allowing it to account for boosted signals predicted by exotic models. Though this provides more freedom to the models, in the case of a detection it would be necessary to check if the converged model is physically plausible. The normalized spherically averaged power spectra of all simulations used in the training set can be seen in Fig.~\ref{fig:all_ps3d_simu}, showing the large variety of power spectrum shapes.

These power spectra were then used to train a VAE with a two-dimensional latent space, meaning that we want to capture the shape of the 21 cm signal using two parameters. It should be noted that due to the 21 cm signal power spectrum being isotropic (to first order, ignoring peculiar velocities), the spherical power spectrum contains all information about the signal under the assumption of Gaussianity. So it is sufficient to train the VAE on the spherical power spectra, rather than the covariance matrix itself. The VAE has two components, an encoder that maps the normalized power spectra to the latent space, and a decoder that can be used to recover the normalized power spectra corresponding to any point in the two-dimensional latent space. Both the encoder and decoder were trained, and the optimization was performed by minimizing the reconstruction loss between the training power spectrum used as input to the encoder, and the recovered power spectrum given as output by the decoder. We divided the simulated power spectra into a training set of 950 power spectra and a validation set of 50 power spectra. The reconstruction loss, defined as the mean squared error (MSE) between the output and the input power spectra, stabilized after 500 out of a total of 4000 iterations. When comparing the behavior of the reconstruction loss for the training and validation sets, no over-fitting was observed. After training, we also checked the ratio between the input and output. A median value of 1 and rms of 0.1 was observed for both the training and validation sets. This rms is well below what is typically expected in terms of measurement errors with the first-generation detection experiments. The power spectrum obtained from the trained decoder at any given point in the latent space can now be used to calculate the frequency-frequency covariance matrix, thus effectively capturing the covariance of the 21 cm signal from the simulations into two latent space quantities. We also tested training on a higher dimension of the latent space but did not find any improvement in the reconstruction. This is likely due to the sparse sampling of the eight-dimensional parameter space using 1000 simulations and a denser sample with a higher latent space dimension could possibly capture subtler changes in the power spectra. However, we defer that to future analyses with higher sensitivities, where such small effects on the power spectrum will be more important.

\subsection{Covariance model}
The trained VAE kernel serves as the 21 cm covariance ($\textbf{K}_{21}$) with three parameters: the two latent space dimensions $x_1$ and $x_2$ and a scaling factor for the 21 cm signal variance $\sigma^{2}_{21}$. For the foregrounds, we used an analytical covariance model, which is a good description of the spectral structure we see in the data based on multiple trials using different combinations of covariance functions. The form of the functions used in our covariance model can be described using the Matern class functions,
\begin{equation}\label{eq:matern}
\kappa_{\mathrm{Matern}}(\nu_{p},\nu_{q}) = \sigma^{2}\dfrac{2^{1-\eta}}{\Gamma(\eta)}\left(\dfrac{\sqrt{2\eta}r}{l}\right)^{\eta}K_{\eta}\left(\dfrac{\sqrt{2\eta}r}{l}\right).
\end{equation}
Here, $\sigma^{2}$ is the variance, \textit{l} is the frequency coherence scale, $r=|\nu_{p}-\nu_{q}|$ is the frequency separation, $\Gamma$ is the Gamma function and $K_{\eta}$ is the modified Bessel function of the second kind. Different values of $\eta$ correspond to different functional forms that are special cases of the Matern class functions. The GP covariance has a foreground, 21 cm, and noise components (Eqs. \ref{eq:gp_data} and \ref{eq:gp_covariances}). However, we find that it is necessary to use two components to model the foregrounds: an "intrinsic" and a "mode-mixing" component. In addition, an "excess" component was used to account for the excess power seen in the data. The different components used in the GP covariance model are:

\paragraph{Intrinsic foregrounds --- $\textbf{K}_{\mathrm{int}}(l_{\text{int}}, \sigma^{2}_{\text{int}})$:} Diffuse Galactic emission and extragalactic point sources below the confusion limit within the FOV constitute the intrinsic foregrounds after sky model subtraction. These foregrounds are expected to have a very large frequency coherence scale due to the smooth spectrum of the synchrotron emission mechanism. We modeled the covariance of the intrinsic foregrounds using a radial basis function (RBF; obtained by setting $\eta=\infty$ in Eq. \ref{eq:matern}), which yields very smooth models along frequency \citep{mertens2018statistical}. We used a uniform prior $\mathcal{U}(20,40)$~MHz on $l_{\text{int}}$ to capture the very large frequency coherence scale features due to intrinsic foregrounds. We note that a wider prior does not affect the converged value of $l_{\text{int}}$ and the specific range $\mathcal{U}(20,40)$ was chosen to have a narrow prior enclosing the converged value, which sped up the ML-GPR runs for the 100 signal injection tests that were performed on ML-GPR (described in Sect. \ref{sec:robustness_mlgpr}).

\paragraph{Mode-mixing foregrounds --- $\textbf{K}_{\mathrm{mix}}(l_{\text{mix}}, \sigma^{2}_{\text{mix}})$:} Interferometers are chromatic instruments and flat-spectrum sources far away from the phase center till the horizon occupy a region in the $k_{\perp},k_{\parallel}$ space known as the foreground wedge \citep{morales2012four, vedantham2012imaging}. Apart from this, additional frequency modulations can be imparted on the foreground data by the chromatic primary beam, especially near nulls and sidelobes, the instrumental bandpass, and other systematic effects. To account for these effects, we used a mode-mixing foreground component. The covariance model used for this component is an RBF but with a $\mathcal{U}(0.1,0.5)$~MHz prior on $l_{\text{mix}}$ accounting for the smaller frequency scale fluctuations due to the mode mixing. An RBF was chosen here since it has a rapid fall in power at high delay due to the smooth models it yields. This also makes it easier to separate it from the 21 cm signal without signal loss and this has been tested through simulations and signal injection tests.

\paragraph{Excess --- $\textbf{K}_{\mathrm{ex}}(l_{\text{ex}}, \sigma^{2}_{\text{ex}})$:} We find that the data cannot be adequately described by just a foreground and a 21 cm component, and there is additional power in the data with a small coherence scale that is difficult to differentiate from the 21 cm signal. This could be caused by small-scale frequency fluctuations introduced into the data by instrumental effects and RFI. Suboptimal calibration and polarization leakage could also be contributing factors to this additional power. This "excess power" is seen to be well described by an exponential covariance model. An exponential function is obtained by setting $\eta$=0.5 in Eq. \ref{eq:matern}. We used a $\mathcal{U}(0.2,2)$~MHz prior on $l_{\text{ex}}$ for this component. We note that even though this prior range is similar to the prior range for \textit{l} in $\textbf{K}_{\mathrm{mix}}$, an exponential kernel does not have a sharp drop in power at large $k_{\parallel}$ like the RBF, making it considerably more difficult to separate from the 21 cm signal component. Hence, this excess component was not subtracted from the data. This avoids a potential signal loss due to the absorption of the 21 cm signal into this component.

\paragraph{Noise --- $\textbf{K}_{\mathrm{n}}$:} The noise covariance is calculated from the time-differenced Stokes-V image cubes as a proxy for the thermal noise. We find that using a fixed noise covariance is sufficient, and multiplying the noise covariance by a scaling factor does not affect our results since such a scaling factor converges to a value very close to 1.

\begin{figure}
        \includegraphics[width=\hsize]{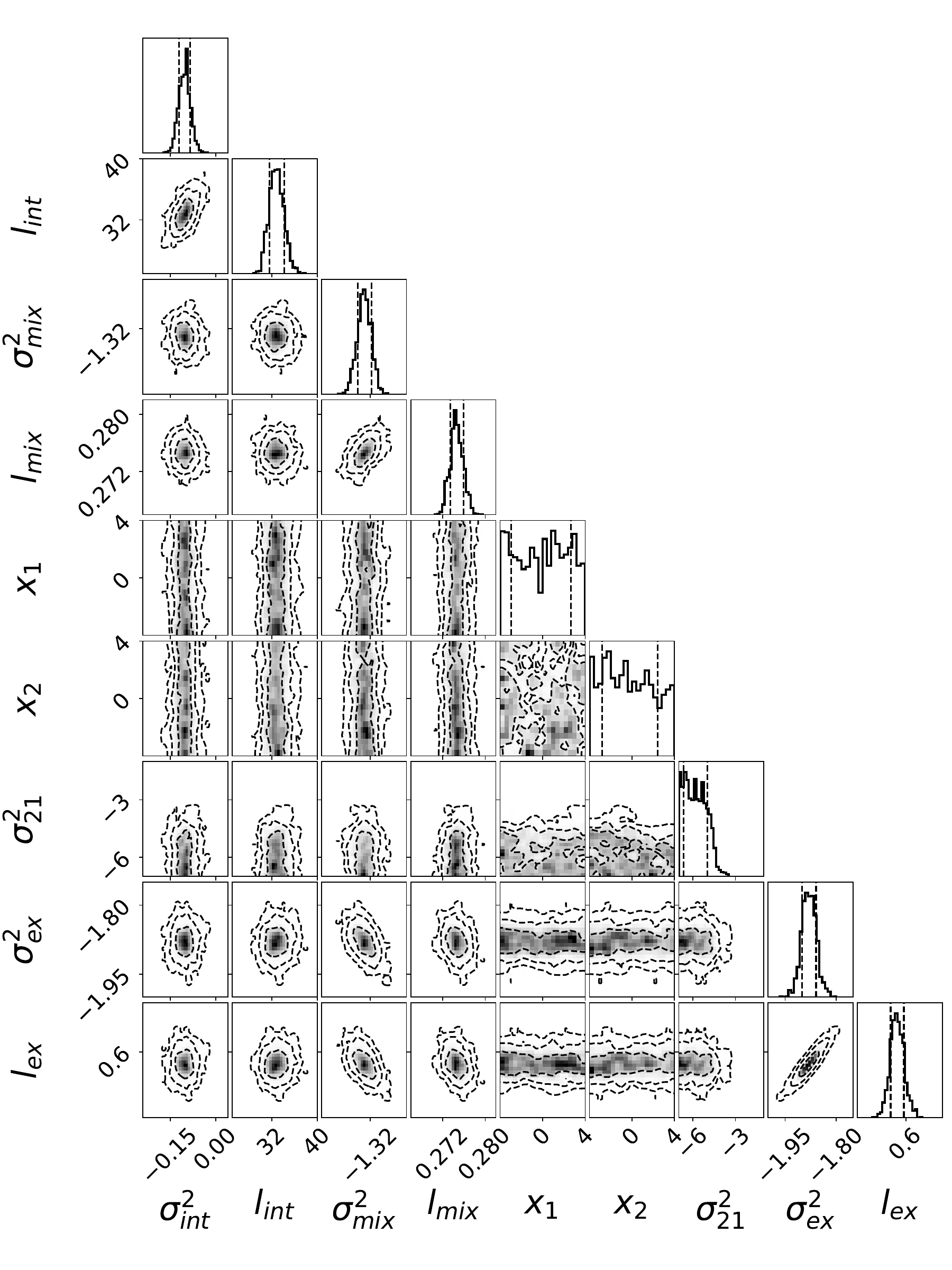}
    \caption{Corner plot showing the posterior probability distributions for the different parameters used in ML-GPR. The dashed contours correspond to the 68\%, 95\%, and 99.7\% confidence levels. The vertical dashed lines in the one-dimensional histogram enclose the central 68\% of the probability.}
    \label{fig:corner}
\end{figure}

\subsection{Application to data}
The data used as input to ML-GPR are the gridded visibility cubes described in Sect. \ref{sec:viscube}. The optimal parameters for our covariance model were obtained using Monte Carlo-based algorithms, Markov chain Monte Carlo (MCMC) or nested sampling, to generate samples from the posterior distribution. Monte Carlo-based methods offer an advantage over simple gradient-based optimization techniques since the former yield both the optimal parameters and the uncertainties associated with them, which can be propagated down to obtain the corresponding uncertainties on the final power spectrum. We get very similar results with both MCMC and nested sampling as the optimization method, but nested sampling, while being more computationally expensive, yields a more complete sampling of the parameter space within the prior bounds and also provides an evidence value. Hence, we used nested sampling with 100 live points to obtain the optimal set of parameters. The prior ranges and converged values for the parameters in our covariance model are listed in Table \ref{tab:mlgpr}. Figure \ref{fig:corner} shows a corner plot of the posterior probability distribution of the parameters. The parameters $x_1$ and $x_2$, which describe the 21 cm signal shape, do not converge, and the variance $\sigma^{2}_{21}$ hits the lower bound of the prior range, as we would expect for data in which the thermal noise level is well above the 21 cm signal. All other parameters for the foreground and excess components converge to well-constrained \textit{l} and $\sigma^{2}$ values, within the prior bounds. Finally, multiple realizations of foreground cubes were sampled from the posterior probability distribution of the parameters and subtracted from the data to yield residual data cubes, which were then used to calculate the residual power spectrum (Fig. \ref{fig:ps_ul}) along with its uncertainties, representing the spread in the distribution of samples in the parameter space.

\begin{figure*}
        \includegraphics[width=\hsize]{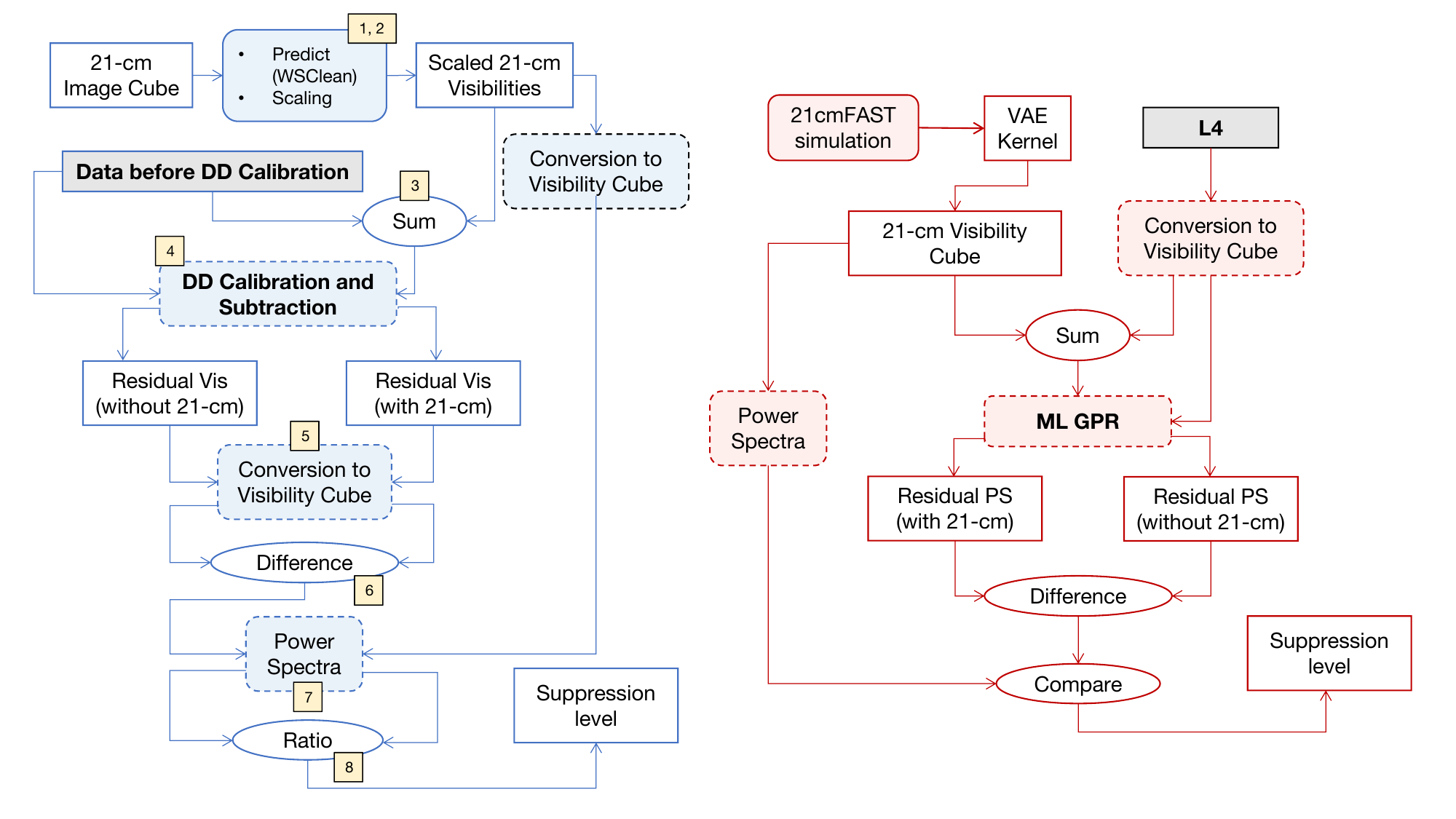}
    \caption{Flowchart for the signal injection tests performed on a DD calibration and subtraction step (outlined in blue), and on ML-GPR (outlined in red). The numbers refer to the sequence of steps described in the text in Sect. \ref{sec:robustness_ddcal}.}
    \label{fig:flowchart_injection}
\end{figure*}

\section{Robustness tests and residual power spectra}\label{sec:robustness}
To ensure that the different calibration and foreground subtraction steps do not result in signal loss \citep[e.g.,][]{patil2016systematic}, it was necessary to perform robustness tests on these steps. The test consists of injecting an artificial 21 cm signal into the data, passing it through the calibration step, and comparing the recovered signal with the injected signal. This is a similar procedure to that followed by \citet{mevius2022numerical}.
The injected signal power is well above the expected 21 cm signal, but as long as it is small compared to the foregrounds and in the linear perturbation regime, the suppression factor is expected to be the same as the actual faint signal \citep{mouri2019quantifying}. We verified this by performing injection tests with different signal strengths. We performed robustness tests on all the steps involving DD calibration and also on ML-GPR. A flowchart describing the workflow of the signal injection simulations is given in Fig. \ref{fig:flowchart_injection}.

\begin{figure*}
        \includegraphics[width=\hsize]{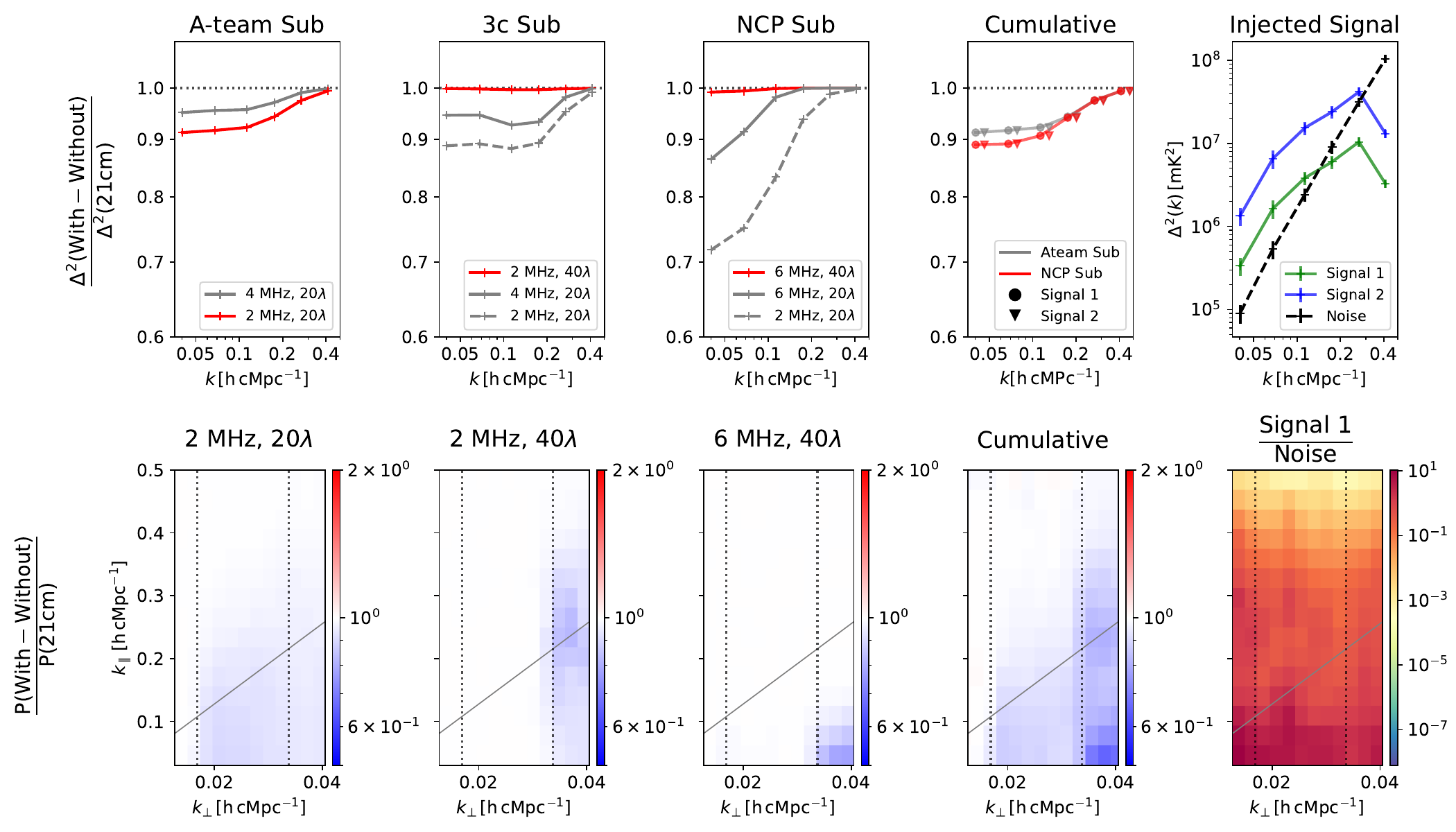}
    \caption{Results of the robustness tests performed on different steps of the calibration pipeline. The results of the tests on the three DD calibration steps for different smoothing kernel widths and baseline cuts are presented in the first three columns from the left. The top row shows the suppression factor in the spherical power spectra for the different steps. The lines in red correspond to the settings that were used for the final analysis, and the plots in the bottom row show the suppression factor in the cylindrical power spectra corresponding to these settings. The vertical dotted lines in the bottom row correspond to the 20$\lambda$ and $40\lambda$ cuts and the region between these two lines is used for constructing the final spherical power spectra. The fourth column from the left shows the cumulative suppression of a signal injected into the L2 data. The top panel shows the cumulative signal suppression after A-team subtraction and after NCP subtraction for the injected 21 cm signal scaled by a factor of 1000 (Signal 1: circles) and 2000 (Signal 2: inverted triangles). The inverted triangle markers have been shifted horizontally for visual clarity. The rightmost column presents the spherical (top) and cylindrical (bottom) power spectra of the injected signal and their comparison with the noise power spectra.}
    \label{fig:robustness}
\end{figure*}

\subsection{Robustness test on DD subtraction steps}\label{sec:robustness_ddcal}
DD calibration of radio interferometric data hinges on the fact that for arrays with a large number of interferometric elements, there are enough excess calibration equations over unknowns to allow us to obtain separate calibration solutions for several directions. However, having a large number of directions also increases the number of free parameters in the calibration model and the risk of over-fitting features other than the DD effects increases. In other words, the MA gains for the different directions can conspire to create a PSF that, convolved with the sky model in that direction, can mimic a part of the 21 cm signal or any other component present in the data and absent in the sky model. This can lead to suppression of the signal and any un-modeled component in the data, in particular on the larger angular and frequency scales \citep{patil2016systematic}. In our analysis, the steps of A-team subtraction, 3C subtraction, and NCP subtraction involve a step of DD calibration followed by model visibility prediction, corruption, and subtraction. Hence, it is necessary to check that these steps do not result in significant signal suppression. Therefore, we performed a signal injection test on all these three calibration steps separately. Below we list the steps followed to perform the signal injection test during the calibration step:

\begin{enumerate}
    \item The mock signal used in this analysis was obtained from the 21 cm brightness temperature cube produced by \citet{jelic2008foreground} using the {\tt 21cmFAST} code, which was also used by \citet{mevius2022numerical}. We note that the signal injection test in the linear perturbation regime is independent of the injected signal \citep{mouri2019quantifying}, and to be self-consistent with earlier work, we chose to use the same cube. This image cube was used to predict the corresponding visibilities employing the {\tt predict} task in {\tt WSClean}, which effectively performs a Fourier transform of the data to generate visibilities at the observed $u \varv w$ coordinates for each frequency. 
    \item The 21 cm simulated visibilities were multiplied by a factor of 1000 to boost up the signal to be close to the thermal noise level of the data for the lowest k modes. At the same time, it was made sure that the injected signal is still more than two orders of magnitude fainter than the foregrounds so that the signal itself has a negligible impact on the amplitude of the gain solutions. 
    \item These scaled simulated 21 cm signal visibilities were added to the data visibilities just before the calibration step being tested. 
    \item The calibration and subtraction were then performed with exactly the same parameter settings as was done for the data without the injected signal. 
    \item The DD subtracted visibilities were then gridded in the manner described in Sects. \ref{sec:image_cubes} and \ref{sec:viscube} to obtain the data cube in the $u\varv\nu$ space. 
    \item The data cube without the injected signal was subtracted from the data cube with the injected signal to obtain the residual 21 cm signal data cube. 
    \item The cylindrical and spherical power spectra for this residual data cube were obtained in the manner described in Sect. \ref{sec:2d1dps}. 
    \item The ratio of these power spectra over the power spectra of the injected 21 cm signal gives the factor by which the signal power is suppressed at the different wave modes due to this particular DD calibration and subtraction step. 
\end{enumerate}
The workflow of the signal injection tests on DD calibration and subtraction is shown on the left of Fig. \ref{fig:flowchart_injection}. The numbers indicate the sequence of steps.

The signal injection test was repeated for different spectral gain smoothing kernel widths and baseline cuts. Smoothness constraints on the gain solutions can decrease signal suppression since it decreases the degrees of freedom of the calibration algorithm by imposing constraints on the parameters that are being solved for, thereby decreasing the chance of over-fitting \citep{barry2016calibration,ewall2016first,patil2016systematic}. Applying a baseline cut involves the use of only the baselines longer than a limit for calculating the calibration solutions. The station-based gain solutions are then applied to the predicted model visibilities and subtracted from all baselines. This restricts any absorption of the signal, and hence signal suppression, to only the baseline range higher than the cut that was used in calibration. Now only the baselines smaller than the cut can be used for power spectrum estimation and there is a much lower risk of signal suppression in these baselines. However, using a baseline cut decreases the accuracy of the gain solutions for a given station, since a small number of baselines that are longer than the cut are available for a given station for calculating the gains. This problem is particularly important for NenuFAR since having a baseline cut larger than the core diameter (400 m) would mean that for a given core station, only three baselines involving the remote stations are available for obtaining the gain solution (for the current NenuFAR configuration with three remote MAs). This decreases the accuracy of the model subtraction and introduces an excess variance on baselines lower than the cut. Also, the baseline-based gains (combination of the gains of two stations) are only constrained by the sky model on long baselines and not so much on short baselines. Hence, any inaccuracy of the sky model and the gains on the longer baselines can lead to spurious signals when applied to the model and subtracted from the short baselines. This can also contribute to excess variance.

The results of the signal injection tests done on the different DD calibration steps are summarized in Fig. \ref{fig:robustness}. As expected, we see that there is negligible signal suppression on the baselines that are not used in the calibration step. Also, using a wider gain smoothing kernel reduces the signal suppression in the baselines that are used in calibration. The final values of the gain smoothing kernel width and baseline cut used for the three different DD calibration and subtraction tests were decided by comparing the results of the signal injection tests with the level of residual power after subtraction.

For the A-team subtraction step, using a 40$\lambda$ baseline cut is not feasible. This is because it is the first step in our calibration pipeline, and nonoptimal A-team subtraction due to the lower number of baselines used in calibration adversely affects all the following steps of DI correction and sky model construction. The smoothing kernel of 2~MHz for A-team subtraction is necessary since a 4~MHz smoothing is not able to account for the small-scale frequency fluctuations of the primary beam at the location of the A-team sources. Using a 4~MHz kernel does decrease the signal suppression factor but results in high residuals at the location of the A-team, which again affects all the successive calibration steps. So we used a frequency smoothing kernel of 2~MHz and a baseline cut of 20$\lambda$ for the A-team subtraction step, thus favoring accurate A-team subtraction and allowing for a < 10\% signal loss for which a bias correction was performed when calculating the power spectra (Sect. \ref{sec:results}).

For 3C subtraction and NCP subtraction, we find that using a baseline cut of 40$\lambda$ does result in higher residual power after subtraction compared to when we used a 20$\lambda$ cut. However, we get negligible signal suppression for a 40$\lambda$ cut and the increase in residual power is less than the inverse of the signal suppression factor for a 20$\lambda$ cut. Thus, we used a 40$\lambda$ cut for both the 3C subtraction and NCP subtraction steps. A frequency smoothing kernel of 2~MHz was used for the 3C subtraction step since it leads to a more accurate subtraction of the 3C sources, which are far from the phase center. For the NCP subtraction step, we used a 6~MHz smoothing kernel since it leads to better source subtraction than a 2~MHz kernel. This is because the phase center is at the NCP and a 6~MHz kernel is sufficient to account for the frequency behavior of the main lobe of the primary beam. Using a 6~MHz kernel also increases the S/N of the gain solutions and we find that the noise in the gain amplitude and phase solutions is smaller when a 6~MHz kernel is used for NCP subtraction. (Appendix \ref{sec:dd_errors}). The S/N becomes an important issue for the NCP subtraction since the total flux is divided into multiple clusters and it is necessary to make sure that the faintest cluster has enough S/N to yield reliable calibration solutions.

\subsection{Robustness test on the full calibration pipeline}\label{sec:robustness_full}
Once the results of the robustness tests for each DD subtraction step were assessed and the final settings to use in each processing and analysis step were selected, we performed a signal injection test on the entire calibration pipeline. This involved adding the scaled 21 cm simulated visibilities to the L2 data and passing it through all the steps of A-team subtraction and DI correction, 3C subtraction, and NCP subtraction one by one. The residual data cube without the injected signal was subtracted from the residual data cube with the injected signal to obtain the residual 21 cm signal data cube. The ratio of the power spectrum of this data cube with that of the injected signal then yields the suppression factor for the entire calibration pipeline. We note that the signal was injected before DI correction, so the calculated DI gains get applied to the injected signal in the course of passing the data through the pipeline. So for all steps after DI correction, to calculate the suppression factor, it is necessary to divide the power spectrum of the residual signal by the power spectrum of the injected signal with these DI gains applied. 

The fourth column in Fig. \ref{fig:robustness} shows the results of injecting the signal to the L2 data and passing it through the pipeline with the final settings. We find that in addition to the A-team subtraction step, the DI correction step introduces a small additional suppression (2\% of the signal). In the top panel, the additional suppression after the A-team subtraction step till the NCP subtraction step comes almost solely from the DI correction step. This additional factor is due to the fact that the calculated DI gains in the direction of the NCP are slightly higher when there is an injected signal. We repeated this signal injection test with twice the magnitude of the injected signal (Signal 2 in the top row, fourth and fifth columns of Fig. \ref{fig:ps2d}). We find that the signal suppression percentage in all DD calibration steps and in the DI correction step remains the same. This is in agreement with the results of \citet{mouri2019quantifying} who conclude that in the linear regime where the signal is much weaker than the foregrounds, the suppression factor due to calibration is independent of the injected signal strength. In the bottom row of Fig. \ref{fig:robustness}, we can see the factor by which the signal will be suppressed in the $k_{\perp},k_{\parallel}$ space due to the calibration pipeline. It should be noted that we only used the baseline range 20$\lambda$ to 40$\lambda$ (the region between the vertical dotted lines) to construct the final spherical power spectra.

\subsection{Robustness test on ML-GPR}\label{sec:robustness_mlgpr}
Since we applied ML-GPR to the gridded visibility cubes, an additional signal injection test was done on the gridded data cube after NCP subtraction. This allows us to test the robustness of the ML-GPR step using a large variety of simulated 21 cm signals. This was not necessary for the calibration steps since the exact shape of the signal does not affect the signal injection results during calibration, as long as the injected signal is small compared to the foregrounds \citep{mouri2019quantifying}. The decoder of the trained VAE kernel (described in Sect. \ref{sec:mlgpr}) was used to generate power spectra corresponding to uniformly spaced points in the two-dimensional latent space. We chose five points between the values $-2$ and 2 for both $x_1$ and $x_2$, making a total of 25 different shapes of the 21 cm power spectrum. For each such signal, the power was scaled to be equal to 0.5, 1, 4, and 20 times the thermal noise power, thus covering a large range of intensities of the 21 cm signal for which we want to test the performance of ML-GPR. Each of these 100 mock 21 cm signals was separately injected into the data. To do this, one realization of the visibility cube was predicted per signal and added to the complex gridded data cube before ML-GPR. The data cube with the injected signal was next used to perform GPR with the same priors as was used for the actual data. The residual power spectrum of the data without an injected signal was then subtracted from the residual power spectrum obtained by applying ML-GPR to data with an injected signal. The workflow used in the signal injection tests for ML-GPR is shown on the right of Fig. \ref{fig:flowchart_injection}. The recovered 21 cm power spectrum from the injection test ($\Delta^{2}_{\text{rec}}$) can be compared to the power spectrum of the injected signal ($\Delta^{2}_{\text{inj}}$) to see if the signal is absorbed in the procedure. We used two metrics to quantify the results of the signal injection simulations, the z-score and bias, defined as follows:
\begin{align}
&\text{z-score(}k\text{)} = \dfrac{\Delta^{2}_{\text{rec}}(k)-\Delta^{2}_{\text{inj}}(k)}{\sigma_{\Delta^{2}_{\text{rec}}}(k)},\nonumber\\
&\text{Bias}(k) = \dfrac{\Delta^{2}_{\text{rec}}(k)}{\Delta^{2}_{\text{inj}}(k)}.
\end{align}

\begin{figure}
    \includegraphics[width=\hsize]{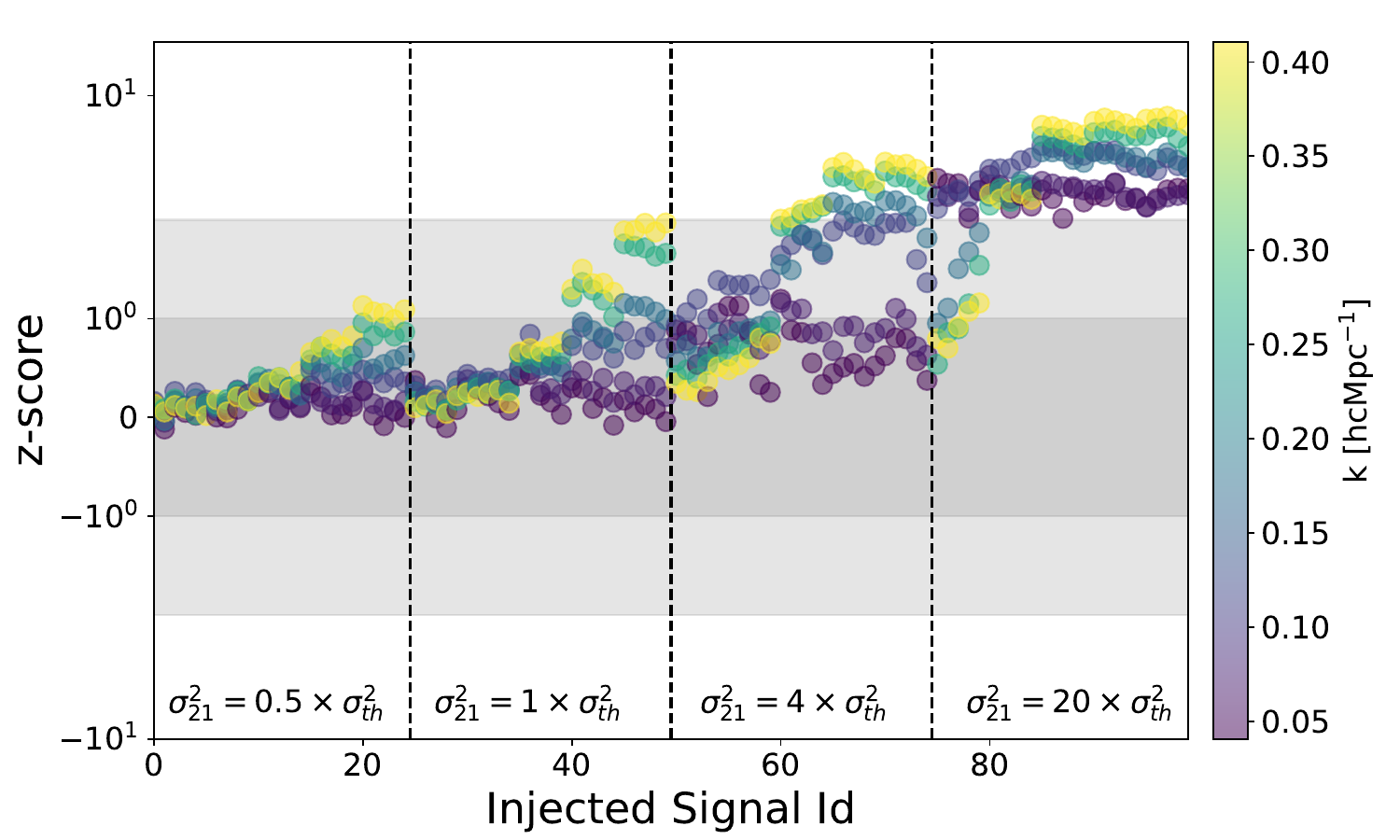}
    
    \includegraphics[width=\hsize]{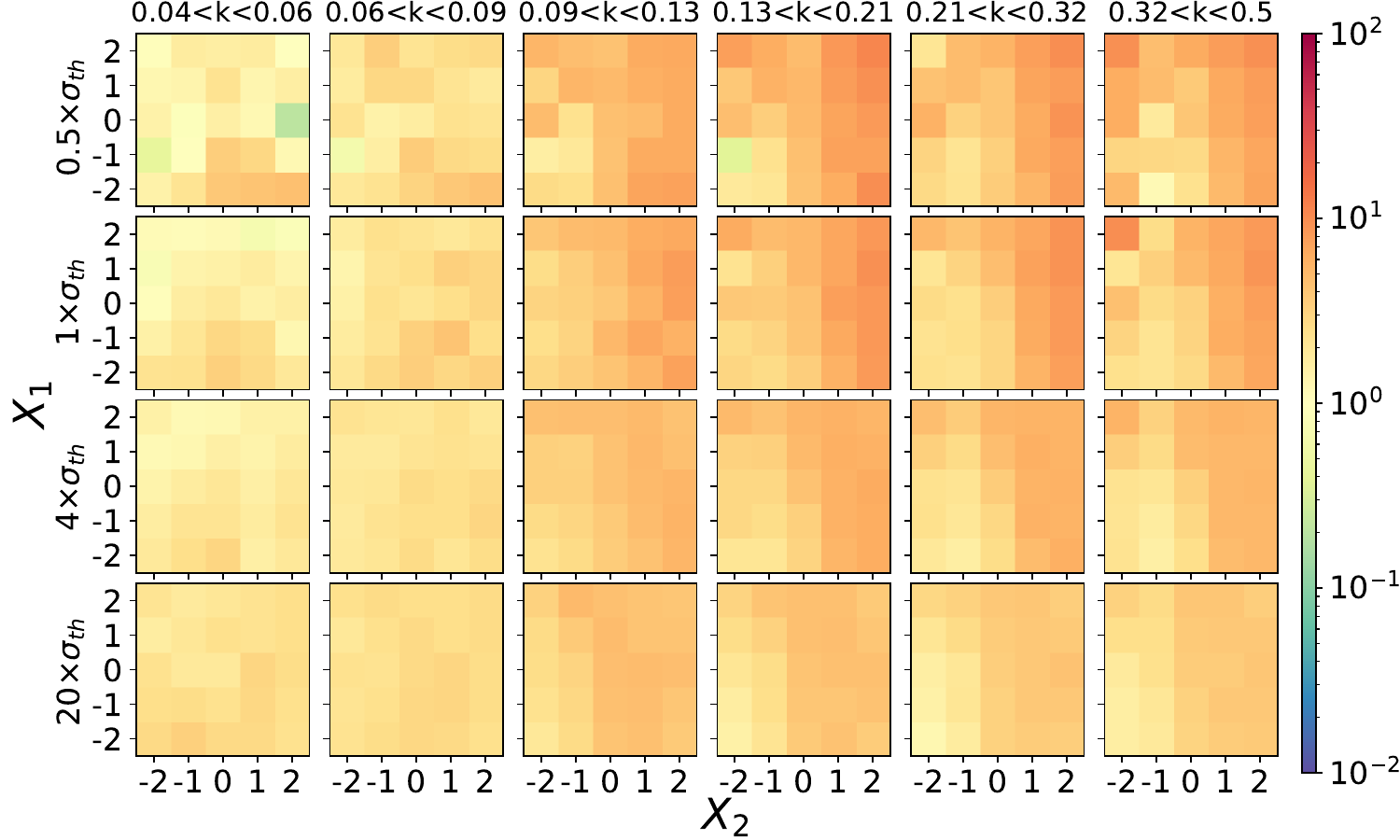}
    \caption{Results of the robustness tests performed on the ML-GPR step. Top: z-score for all injected signals at the different $k$ bins. The vertical dashed lines demarcate the different strengths of the injected signals. The horizontal gray bands indicate the 1-sigma and 2-sigma levels. Bottom: Bias for all injected signals at the different $k$ bins. The 25 cells in each panel correspond to the 25 different shapes of the injected signal. The different rows correspond to the different strengths of the injected signal and the different columns show the results in different $k$ bins. The color scale indicates the value of the bias in each case.}
    \label{fig:zscores_bias}
\end{figure}

The z-score specifies the number of standard deviations by which the recovered signal power spectrum over or underestimates the input. A negative z-score indicates suppression of the signal and a value below $-2$ would thus indicate a suppression of the signal beyond the 2-sigma upper limits at that particular $k$ bin. The z-scores for all signal injection tests are shown in the top panel of Fig. \ref{fig:zscores_bias}. The different $k$ bins are indicated with different colored points. We find that none of the signals are suppressed beyond the 1-sigma error bars, and thus the 2-sigma upper limits are comfortably valid. The z-score for brighter signals has a large positive value since $\sigma_{\Delta^{2}_{\text{rec}}}(k)$ have very low values compared to the signal strength for stronger signals. This is because the contribution to the error bar from the uncertainty in the converged hyperparameter values is lower for stronger signals since the signals are estimated with higher precision. The bias tells us the factor by which the recovered signal power is higher than the injected signal. The bias at each $k$ bin for all signal injection tests is illustrated in the bottom panel of Fig. \ref{fig:zscores_bias}.  The signals for which the bias values are less than unity are all within the 1 sigma limits and correspond to the points in the top panel that lie below zero. There is a positive bias, particularly for signals with high $x_2$ values. This suggests that there is some degeneracy between the foregrounds, excess, and the 21 cm components when modeled as Gaussian processes, but it only results in the absorption of the foregrounds into the excess or the 21 cm signal component, which are not subtracted from the data. This is acceptable as long as we set upper limits on the magnitude of the signal, and do not claim a detection.

\begin{figure*}
        \includegraphics[width=\hsize]{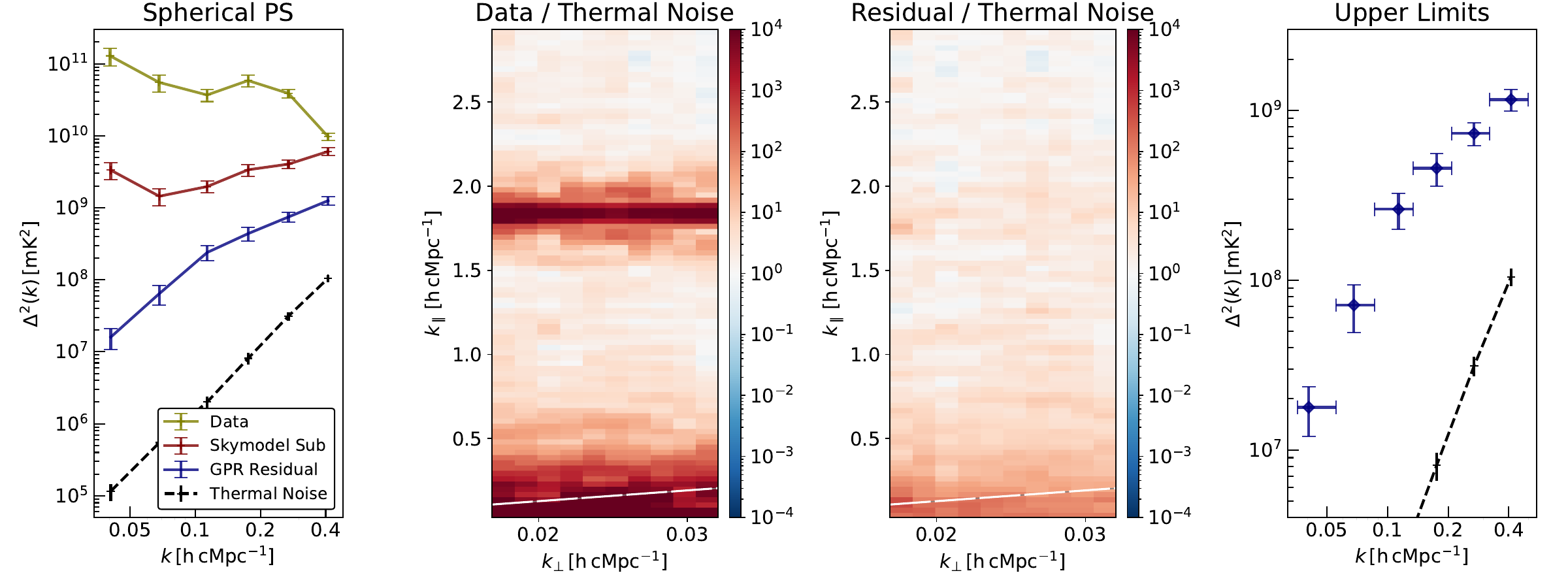}
    \caption{Spherical and cylindrical power spectra at some key stages of the analysis pipeline. The left-most panel shows the spherical power spectra after preprocessing ("Data"), after sky model subtraction ("Skymodel Sub"), and after GPR ("GPR Residual"), along with the thermal noise power spectrum ("Thermal Noise"). For the cylindrical power spectra (second and third panels), the ratio with respect to the thermal noise power spectrum is shown. The white dashed lines indicate the horizon limit. The spherical power spectra after noise bias subtraction and suppression factor correction are shown in the rightmost panel.}
    \label{fig:ps_ul}
\end{figure*}

\subsection{Residual power spectra}\label{sec:results}
The residual foreground subtracted data cube from ML-GPR was used to compute the spherical power spectra in logarithmically spaced bins in $k$ space between $k=0.035$ $h\, \textrm{cMpc}^{-1}$ and $k=0.5$ $h\, \textrm{cMpc}^{-1}$. To compute the spherical power spectra, we used only the data in the $u\varv$ range of $20\lambda-40\lambda$, in order to avoid the signal suppression in calibration due to the 3C subtraction and NCP subtraction steps. The spherical and cylindrical power spectra before and after the entire processing are shown in Fig. \ref{fig:ps_ul}. We find that the residual power spectra are around two orders of magnitude higher than the thermal noise level at the lowest $k$ bins, contributed mainly by the strong excess component that is left in the data after GPR. We note that the thermal noise power spectrum for "Residual" is slightly lower than that for "Data" since the post-calibration RFI flagging removes 7.5\% of the data.

\begin{table}[b]
\caption{ 2$\sigma$ upper limits on the 21 cm power spectrum $\Delta^{2}_{ul}(k)$ at different $k$ bins. $\Delta^{2}_{21}(k)$ is the residual power, $\Delta^{2}_{21, err}(k)$ is the error on the residual power, and $\Delta^{2}_{N}(k)$ is the thermal noise power.}
\label{tab:ul}
\centering
\begin{tabular}{@{}ccccc@{}}
\toprule
k bin                 & $\Delta^{2}_{21}(k)$   & $\Delta^{2}_{21, err}(k)$ & $\Delta^{2}_{ul}(k)$   & $\Delta^{2}_{N}(k)$    \\
{(}$h\, \mathrm{cMpc}^{-1}${)} & {(}$\mathrm{K}^{2}${)} & {(}$\mathrm{K}^{2}${)}    & {(}$\mathrm{K}^{2}${)} & {(}$\mathrm{K}^{2}${)} \\ \midrule
0.036 -- 0.055                         & 17.81                          & 2.90                           & 23.60                          & 0.12                           \\
0.055 -- 0.086                         & 71.30                          & 11.11                           & 93.52                          & 0.55                           \\
0.086 -- 0.134                         & 261.63                         & 30.96                          & 323.54                         & 2.01                           \\
0.134 -- 0.207                         & 456.33                         & 49.54                          & 555.41                         & 8.10                           \\
0.207 -- 0.322                         & 731.75                         & 57.18                          & 846.12                         & 31.29                          \\
0.322 -- 0.5                           & 1157.37                        & 85.06                          & 1327.49                        & 104.66                         \\ \bottomrule
\end{tabular}
\end{table}

The noise bias at each $k$ bin was calculated by passing the time-differenced Stokes-V image cube (see Sect. \ref{sec:image_cubes}) through the power spectrum pipeline, and accounting for the factor of 2. This was subtracted from the residual power spectrum to obtain the noise bias subtracted dimensionless power spectrum. In Sect. \ref{sec:robustness_full}, we found that there is a maximum of 11\% suppression in the signal as a cumulative effect of the entire calibration pipeline, coming mainly from the A-team subtraction and DI correction steps in which the $20\lambda$ to $40\lambda$ baselines were used (Fig. \ref{fig:robustness}). To account for this, we performed a bias correction, which amounts to applying the inverse of the calculated average cumulative signal suppression factors at each $k$ bin to the residual power spectra. We place the upper limits at a level of 2$\sigma$ above these values at the different $k$ modes. The power spectrum values with the 2$\sigma$ error bars after noise bias subtraction and after bias correction are shown in Fig. \ref{fig:ps_ul} in the rightmost panel. The 2$\sigma$ upper limits are listed in Table \ref{tab:ul}. The best 2$\sigma$ upper limit is $2.4 \times 10^7 \textrm{mK}^{2}$ at $k=0.041$ $h\, \mathrm{cMpc}^{-1}$.

\section{Discussion and next steps}\label{sec:discussion}
In this section, we discuss the results and the limitations of the current analysis and present the plans for future improvements to the analysis pipeline.
\subsection{Nature of the "excess"}\label{sec:nature_excess}
The residual data after foreground subtraction is seen to be dominated by excess power up to two orders of magnitude beyond the thermal noise limit. This excess power is not possible to subtract from the data using GPR since its spectral behavior is similar to what we expect from the 21 cm signal.

\begin{figure}
        \includegraphics[width=\hsize]{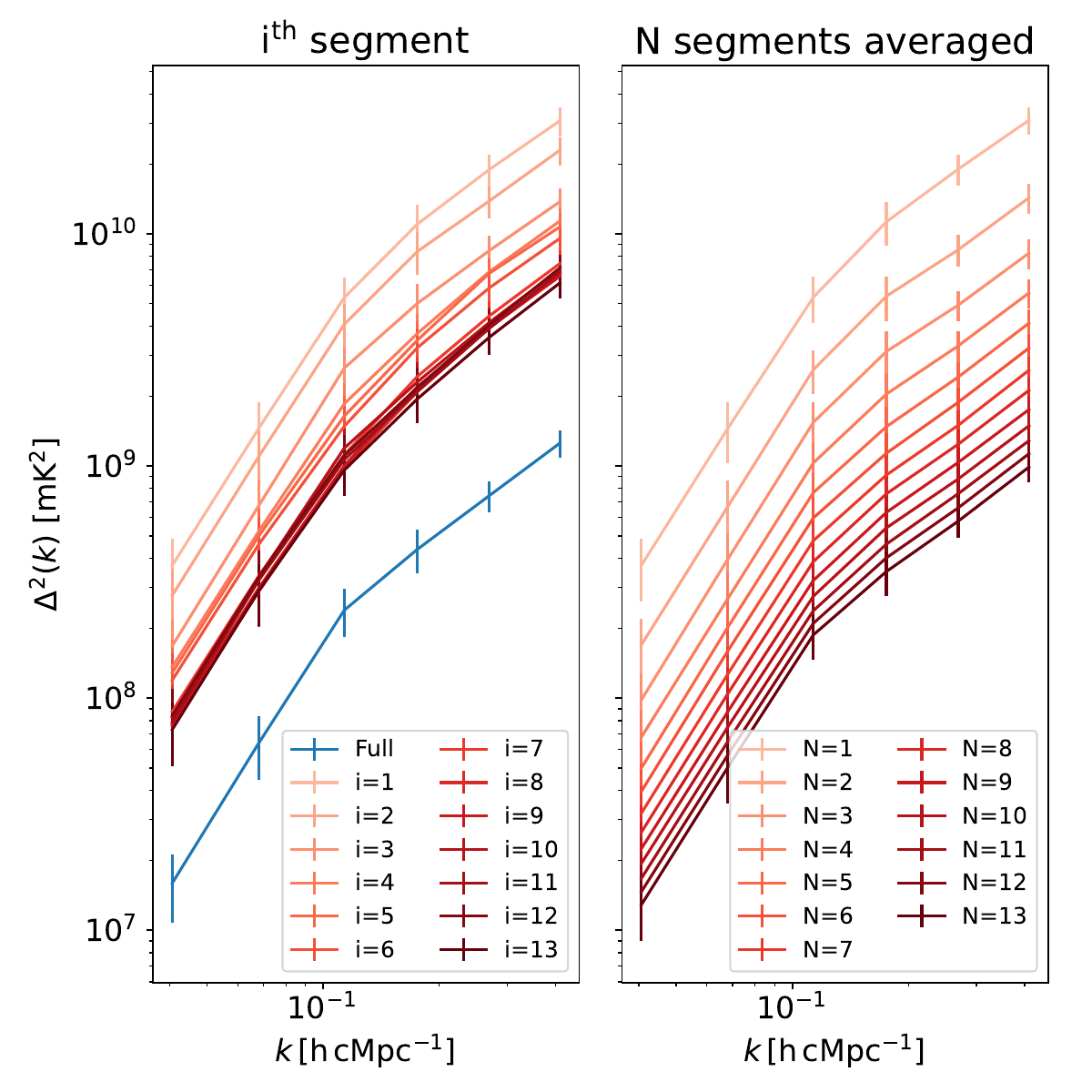}
    \caption{Spherical power spectra of the data after applying GPR on individual time segments. The left panel shows the power spectra of the GPR residuals for individual time segments of the data and the right panel shows the power spectra with increasing volume of data, in the order of observation.}
    \label{fig:residual_segments}
\end{figure}

In order to understand the behavior of the excess as a function of time, we utilized the fact that the visibility data were divided into 13 time segments of 52 min duration each (with a 56 min last segment). The gridded data cube for each of these segments was obtained following Sects. \ref{sec:image_cubes} and \ref{sec:viscube}. A separate GPR was performed on each cube, with the length scales fixed to the values obtained for the full data, and the variances left as free parameters, which were optimized in GPR. Fixing the length scales prevents overfitting of the data, and optimizing the variances allows the time dependence of the strength of the GPR components to be modeled over timescales longer than a time segment. The maximum a posteriori solutions of the foreground cubes were subtracted from the data cube of each time segment to give the residual data cubes, which were used to construct the respective power spectra, shown in the left panel of Fig. \ref{fig:residual_segments}. The excess is stronger in the first six segments, and the last seven segments have a similar power level. We note that the last segment has the lowest power because it is 56 min long and has a lower thermal noise power level than the other 52 min segments.

Possible contributors to this excess are the residuals from Cyg A, which dominate the $u\varv$ plane after sky model subtraction as seen in Fig. \ref{fig:uvfeatures}. Also, Cyg A goes to lower elevations as time progresses and the beam gain in its direction has very low values after the first 6 segments (Fig. \ref{fig:ateams_altitude}). The residuals from Cyg A subtraction are expected to be directly related to its apparent brightness and this is confirmed by wide-field images of the data (for example, the images on the left panels of Fig. \ref{fig:clean_images}) where the residuals at its location are seen to go down with time. Thus, Cyg A residuals are expected to be a major contributor to the excess. It is important to note that this excess power is a result of the variance on the NCP field caused by the residual PSF sidelobes of Cyg A. Therefore, even if the peak flux at the position of the Cyg A is reduced to be below the thermal noise limit, it does not necessarily remove the effect of its PSF sidelobes on the NCP field where the power spectrum is computed. This can only be improved in the future with careful time and frequency modeling of the beam, ionosphere, and any other DD effects, during calibration.

To determine if the excess power is coherent with time within a single night, the power spectrum was constructed with data from an increasing number of time segments included. This was done by computing the weighted average (using the $u\varv$ weights) of the residual data cubes for those segments and constructing the power spectra. It should be noted that the weighted average of all the time segments retrieves the exact full data cube since the gridding is linear. The residual power is seen to go down as we integrate more data (right panel of Fig. \ref{fig:residual_segments}). We find that the power spectrum indicated by "N=13" has slightly lower power than that of "full data." This is because, in the case of N=13, GPR has been applied to the individual data cubes with the variances of the individual GPR components allowed to vary in time instead of having a single variance representing the strength of the GPR components for the duration of the entire observation as is the case for full data, thus allowing slightly better foreground modeling and subtraction. The ratio between the power of the first segment and that of the full data is seen to be more than a factor of 20, as opposed to a factor of 13 that we expect from using 13 times as much data if the excess is incoherent between time segments. This is likely due to the first few segments having more power from the Cyg A residuals, as discussed above. The fact that the power goes down by a factor of 13 or more as we integrate 13 times as much data suggests that the excess power, which dominates the residuals, is incoherent. Thus, as we integrate more nights of observations, it is possible that this excess could go down significantly with time if a part of the excess is incoherent across different nights. This needs further investigation by processing multiple nights of data. However, even if this is confirmed, the final power spectra would remain well above what can be expected from thermal noise alone, which also averages down in a similar manner.

\subsection{Impact of wide-field sources}
\begin{figure}
        \includegraphics[width=\hsize]{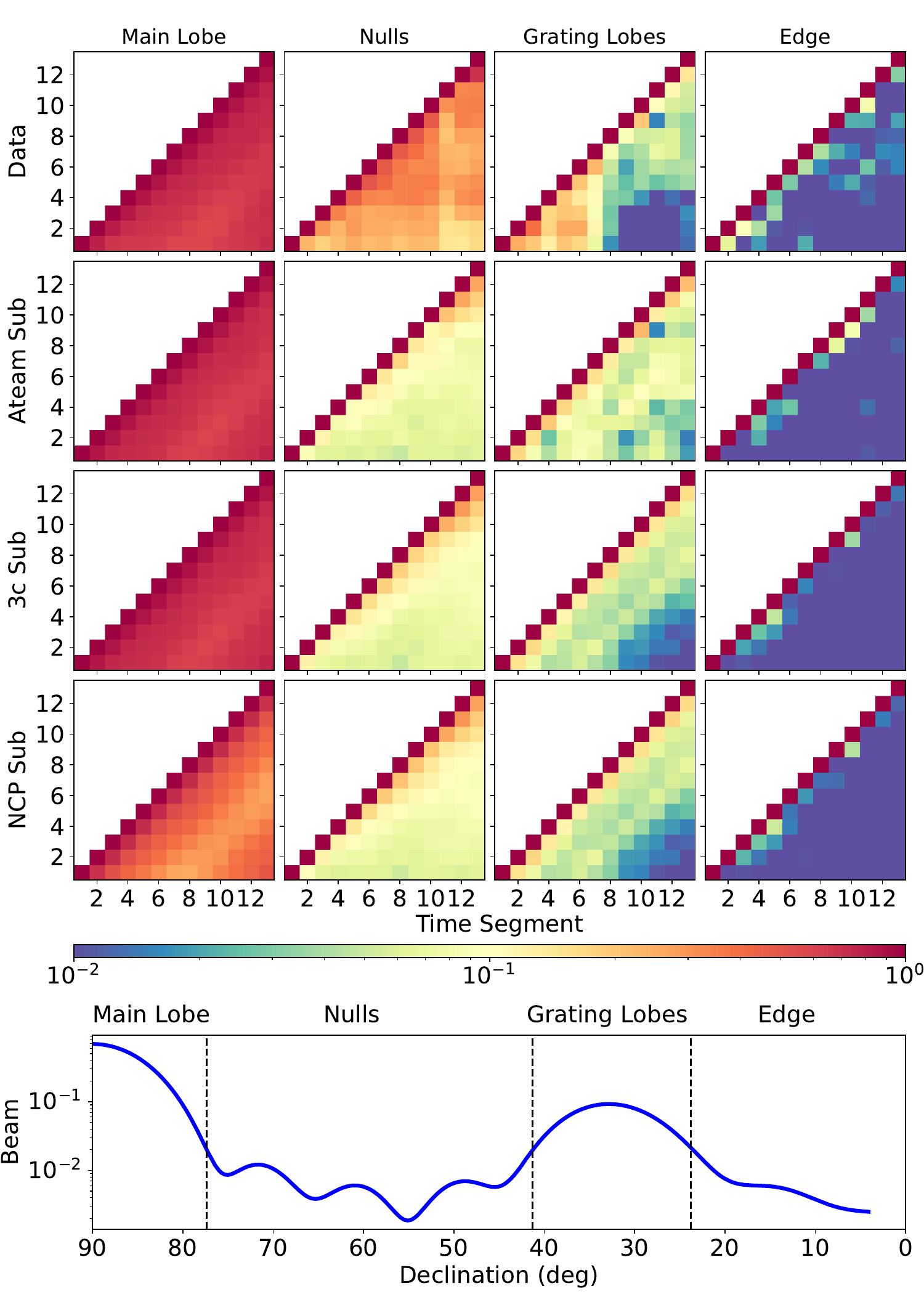}
    \caption{Cross-coherence of the data between different time segments. The top panel shows the cross-coherence computed in four different regions of the primary beam (different columns), for the data at four different stages of sky model subtraction (different rows). The bottom panel shows what the four regions correspond to in the radial profile of the time-averaged primary beam.}
    \label{fig:cross_coherence}
\end{figure}

The NenuFAR primary beam has very strong grating lobes roughly 60 degrees away from the phase center. Wide-field images of the data reveal a large number of strong radio sources falling in these grating lobes, which remain in the data after sky model subtraction. In the 3C subtraction step, seven of the brightest among these sources were subtracted through DD calibration. However, the remaining sources have too little S/N to obtain reliable calibration solutions. Though these sources are far away from the phase center, the PSF sidelobes of these sources will still pass through the NCP field and add a significant contribution to the excess power. To understand the impact of these sources far away from the phase center, we computed wide-field MFS dirty images from the data for each of the time segments into which the data had been divided. Each image was divided into four annular regions around the NCP, labeled "Main Lobe," "Nulls," "Grating Lobes," and "Edge," based on the amplitude of the radial profile of the NenuFAR primary beam (bottom panel in Fig. \ref{fig:cross_coherence}). For the pixels within the $n^{\text{th}}$ annulus, the cross-coherence between the $i^{\text{th}}$ and $j^{\text{th}}$ time segments is given by
\begin{equation}
C_{ij}(n) = \dfrac{\langle I_{i}(l,m)I_{j}(l,m)\rangle}{\sqrt{\langle I^{2}_{i}(l,m)\rangle \langle I^{2}_{j}(l,m)\rangle}},
\end{equation}
where $\langle \cdots \rangle$ indicates an average over all \textit{l,m} lying within the $n^{\text{th}}$ annulus. Figure \ref{fig:cross_coherence} shows the cross-coherence at four different calibration and foreground subtraction stages. It should be noted that the cross-coherence of a segment with itself is unity by definition.

Before A-team subtraction, at the "data" level, the nulls annulus is dominated by flux from Cas A, and the grating lobes annulus is dominated by Cyg A. Cas A is high in the sky throughout the observation, but Cyg A goes below the horizon halfway through the observation, resulting in lower cross-coherence values for the later time segments. After A-team subtraction, the grating lobes annulus is dominated by the strong 3C sources. The 3C subtraction step removes seven of the brightest sources, but the remaining sources have a significant cross-coherence in the grating lobes annulus. There is also higher cross-coherence for the separation of five time segments, corresponding to a time difference of 4.3 h. This is possibly because of the six-fold symmetry of the hexagonal MAs, which results in the grating lobes in the sky that overlap exactly once every 4 h. In the main lobe, the cross-coherence values are close to unity until before NCP subtraction. NCP subtraction decreases the cross-coherence in the main lobe and the modulation seen in the cross-coherence with increasing time difference is because there are large residuals from sources at the edge of the main lobe. As the elongated main lobe rotates against the sky, it completes close to a 180 degree rotation in the duration of the observation as a result of which the time segments separated by 13 segments again have high cross-coherence values. We note that this modulation due to the rotation of the non-axisymmetric main lobe of the beam is also seen before NCP subtraction, but the effect is lower since the field has a more uniform distribution of flux before NCP subtraction. The edge region is dominated by noise, so there is negligible cross-coherence between time segments.

These cross-coherence results show that the timescale of coherence is heavily dependent on the distance from the phase center because of the primary beam. For the main lobe, this timescale is nearly the duration of the full night, while for the nulls, the coherence drops to 20\% or less for consecutive segments. For the grating lobes, the coherence is about 10\% for the separation of between one and six segments and falls to less than 2\% for longer timescales. If the edge region is a reasonable estimate of the noise, we can state that the coherence of the nulls and grating lobes (and, of course, the main lobe) are still dominated by real sources and not too much diluted by noise. These sources behave incoherently in their flux or position from segment to segment due to the primary beam fixed to the Earth. Thus, the power from these sources, coupled with the non-smooth spectral nature of the primary beam at the nulls and grating lobes, could have a large contribution to the excess power at high $k_{\parallel}$ seen in the residual data, which averages down with time like incoherent noise (as seen in Sect. \ref{sec:nature_excess}), on timescales that exceed their own coherence time (as seen in Fig. \ref{fig:cross_coherence}).

\subsection{Limitations and future improvement}
This first analysis on a single night of NenuFAR observations reveals the limitations of the current analysis pipeline. Here, we describe these limitations and discuss future improvements in the processing pipeline necessary to overcome them.
\paragraph{Near-field RFI sources:} Near-field imaging (i.e., imaging of sources on the ground plane in the vicinity of the array) of the data after calibration and sky model subtraction revealed that two prominent sources of local RFI contaminate the data. The approach adopted in this analysis to mitigate its impact was by identifying the baselines where these RFI sources have a strong signature in the delay power spectra and flagging those baselines completely. However, this is an overly simplistic approach, and all un-flagged baselines still have some power due to these local RFI sources. It was seen that flagging the affected baselines significantly reduced excess power at high delay modes. So, local RFI in the unflagged baselines could make up much of the excess power we see in the data. A more targeted approach in mitigating the RFI sources could be by filtering them out from delay or fringe rate power spectra of all baselines, utilizing the fact that the RFI sources have a localized signature in the delay-fringe rate space. We defer that to a future analysis. We note that the impact of the near-field RFI sources is expected to be maximum for NCP observations since the NCP is the only point in the northern sky that is fixed with respect to the Earth, and the local stationary RFI sources add up coherently at the NCP even as the sky rotates with time. This is less the case for other fields, not including the NCP, and the contaminating effects of these local RFI sources on observations of these fields could be significantly less.

\paragraph{Impact of diffuse (polarized) foregrounds:} NenuFAR, with its dense core consisting of many small baselines, is very sensitive to diffuse Galactic foregrounds. These (polarized) foregrounds are not present in the sky model used in calibration, and the power from the diffuse foregrounds can be absorbed in the station gains, leading to calibration errors. We used a $20\lambda$ baseline cut in calibration for the A-team subtraction step and a $40\lambda$ cut for the other calibration steps. This avoids the power due to diffuse foregrounds at spatial modes measured on baselines shorter than $20\lambda$ and $40\lambda$, respectively. However, using a baseline cut results in excess power on the excluded baselines. Thus, both 3C and NCP subtraction result in an excess noise on the baselines smaller than $40\lambda$, which are used in power spectrum generation. The errors in calibration due to diffuse foregrounds can be improved by including diffuse foreground models. However, the ideal way to model the diffuse foregrounds, perhaps using shapelets decomposition \citep{yatawatta2011shapelets,gehlot2022degree}, must still be tested. A future step could also be to build a broadband spectral model of diffuse foregrounds by combining AARTFAAC High Band Antenna and NenuFAR observations.

\paragraph{Beam model in calibration:} The beam model of NenuFAR is currently not integrated into the calibration algorithm. As a result, we performed calibration against an apparent sky model obtained by multiplying the intrinsic sky model with the average beam for each time segment of data. The effect of the beam on the apparent flux of sources on timescales more than the calibration solution interval is accounted for by the calibration gain solutions. However, the variation of the beam on time and frequency scales shorter than a solution interval is currently not accounted for. The high residuals due to Cyg A (seen in the middle-left panel of Fig. \ref{fig:clean_images} and the middle and right panels of Fig. \ref{fig:uvfeatures}), which are expected to be a major contributor to the excess variance (Sect. \ref{sec:nature_excess}), is partly a consequence of this. The solution interval used in the A-team subtraction step is not small enough to model the primary beam accurately at the grating lobes in which Cyg A lies. This could be addressed to some extent when the primary beam model is integrated into the calibration software, and the level of improvement will depend on the accuracy of the beam model.

For the NCP subtraction step, including a beam model in the calibration step will allow us to use longer solution intervals since this will partly mitigate the requirement of the gain solutions to represent the rotating non-axisymmetric beam. This will lower the expected thermal noise for each solution interval and thus make it possible to use smaller clusters with enough flux to be significantly beyond the thermal noise level. Thus, it will be possible to account for DD effects on smaller spatial scales than currently possible. Including a beam model in the calibration software will also mean that the gain solutions do not need to account for the frequency behavior of the primary beam. Therefore, even with a large value for the smoothing scale, it will be possible to fit and subtract the sources well, thus avoiding increasing the risk of overfitting by decreasing the width of the frequency smoothing kernel.

Another advantage of having the beam in the calibration step is using a wider sky model of the NCP field. This is because the spatial gradient of the source fluxes due to attenuation by the beam will now be partly accounted for by the beam model, thus allowing us to use wider clusters with higher S/N. We see in the bottom-right panel of Fig. \ref{fig:clean_images} and the bottom-left panel of Fig. \ref{fig:cross_coherence} that after NCP subtraction, there are residual sources near the edge of the main lobe of the primary beam. Using an extended sky model, along with the beam model, will allow us to subtract these sources better.

Finally, including a beam model will allow us to subtract many sources in the grating lobes of the NenuFAR primary beam. Currently, these sources cannot be subtracted because to model the beam, we needed to calibrate and absorb the beam in the gain solutions. Performing such a calibration run is impossible since the sources do not have enough S/N to yield reliable calibration solutions within the time and frequency solution interval needed to represent the beam faithfully. Once a beam model is available, the visibilities for these sources with the primary beam attenuation can be predicted and subtracted directly from the data. The accuracy of this subtraction will depend on the accuracy of the simulated beam model. We note that all sources that are not subtracted and the residuals from the subtracted sources will continue to have the effect of the rotating beam and will still contribute to the excess power.

\paragraph{Confusion noise limit:} The NCP subtraction step is limited by confusion noise due to the maximum angular resolution of NenuFAR, which is given by the baseline formed by two remote MAs. Once more remote MAs are available, a 5 km maximum baseline in the future will improve the angular resolution by a factor of $\approx 3.4$ and allow us to construct a deeper sky model due to a lower confusion limit (by a factor of $\approx 6.7$). This will allow us to use smaller clusters in the NCP subtraction step since smaller clusters will have a high enough S/N to give reliable calibration solutions.

\paragraph{Polarization leakage:} In this analysis, we have used {\tt diagonal} mode in calibration with {\tt DDECal} and Stokes-I and Stokes-V data for all steps. However, we expect some polarization leakage in NenuFAR since it is a wide-field phased array instrument. Additionally, due to its dense core, NenuFAR is particularly sensitive to diffuse foregrounds, which are expected to have a polarized component. The leakage of Faraday rotated polarized foregrounds into Stokes-I can result in the foregrounds assuming a non-smooth spectral behavior. This can be tested with simulations once the full Stokes beam model is available, and the inclusion of such a beam model in calibration can potentially decrease instrumental polarization.

\paragraph{Ionospheric effects:}
An important effect that can impact the power spectrum is ionospheric scintillation noise \citep{vedantham2015scintillation,vedantham2016scintillation,gehlot2018wide}. A turbulent ionosphere introduces phase shifts on the incoming electromagnetic waves, which can cause the sources to move around in the sky with time, thus affecting the accuracy of their subtraction through calibration. In this analysis, we have selected a relatively quiet ionospheric night, and the calibration solution time intervals used at each stage are longer than the expected timescales of ionospheric effects, making it impossible to solve for these effects. A detailed simulation of the ionosphere is necessary to quantify its impact on NenuFAR data and will be done in the future.\\

In summary, a few specific effects could be the dominant contributors to the excess power we see in the data. Targeted improvements in the processing pipeline can account for, mitigate, and possibly eliminate these effects.

\section{Summary}\label{sec:summary}
This paper describes the first end-to-end analysis of the spectral window of 61 to 72~MHz from an 11.4~h observation of the NCP field with NenuFAR. We draw the following main conclusions from this analysis:

\begin{itemize}
    \item We set the best 2$\sigma$ upper limits on the 21 cm power spectrum at $2.4 \times 10^7 \textrm{mK}^{2}$ at $k=0.041$ $h\, \mathrm{cMpc}^{-1}$ and $z=20.3$. These upper limits are two orders of magnitude higher than the thermal noise power spectrum.
    \item Localized RFI sources strongly impact the data and are a major contributor to the excess power. The RFI is mitigated to an extent by baseline flagging, and a more targeted approach to filtering these local RFI sources could significantly mitigate the excess power.
    \item The excess power seems largely incoherent with time within a single night and averages down like noise as we include more data to construct the power spectra. Processing of multiple nights is necessary to understand whether coherent averaging of multiple nights of data will decrease this excess power.
    \item Many sources lying in the nulls and grating lobes of the NenuFAR primary beam are left in the data after sky model subtraction. Coupled with the frequency fluctuations of the primary beam in these regions, these sources could strongly contribute to the excess power seen in the data. Including a simulated primary beam model in the calibration step would allow us to subtract these sources to some extent and partly mitigate this effect.
\end{itemize}
We will continue to investigate the origin of this excess power and implement improvements in the analysis pipeline to mitigate it. This will allow us to exploit the full sensitivity of NenuFAR. In parallel, we plan to process multiple nights of observations and obtain upper limits closer to the thermal noise and expected 21 cm signal levels, which will allow us to constrain the underlying astrophysical models.

\begin{acknowledgements}
SM, LVEK, and BKG acknowledge the financial support from the European Research Council (ERC) under the European Union’s Horizon 2020 research and innovation programme (Grant agreement No. 884760, "CoDEX”). FGM acknowledges the support of the PSL Fellowship. AB acknowledges support from the European Research Council through the Advanced Grant MIST (FP7/2017-2022, No. 742719). EC would like to acknowledge support from the Centre for Data Science and Systems Complexity (DSSC), Faculty of Science and Engineering at the University of Groningen. AF is supported by the Royal Society University Research Fellowship. The post-doctoral contract of IH was funded by Sorbonne Université in the framework of the Initiative Physique des Infinis (IDEX SUPER). We are grateful to Bradley Greig for his valuable input and guidance in setting up {\tt 21cmFAST} simulations for cosmic dawn. This paper is based on data obtained using the NenuFAR radiotelescope. NenuFAR has benefitted from the following funding sources : CNRS-INSU, Observatoire de Paris, Station de Radioastronomie de Nançay, Observatoire des Sciences de l'Univers de la Région Centre, Région Centre-Val de Loire, Université d'Orléans, DIM-ACAV and DIM-ACAV+ de la Région Ile de France, Agence Nationale de la Recherche. We acknowledge the Nançay Data Center resources used for data reduction and storage.
\end{acknowledgements}

\bibliographystyle{aa}
\bibliography{aa}

\begin{appendix}
\section{RFI statistics}\label{sec:rfi}
\begin{figure}
        \includegraphics[width=\hsize]{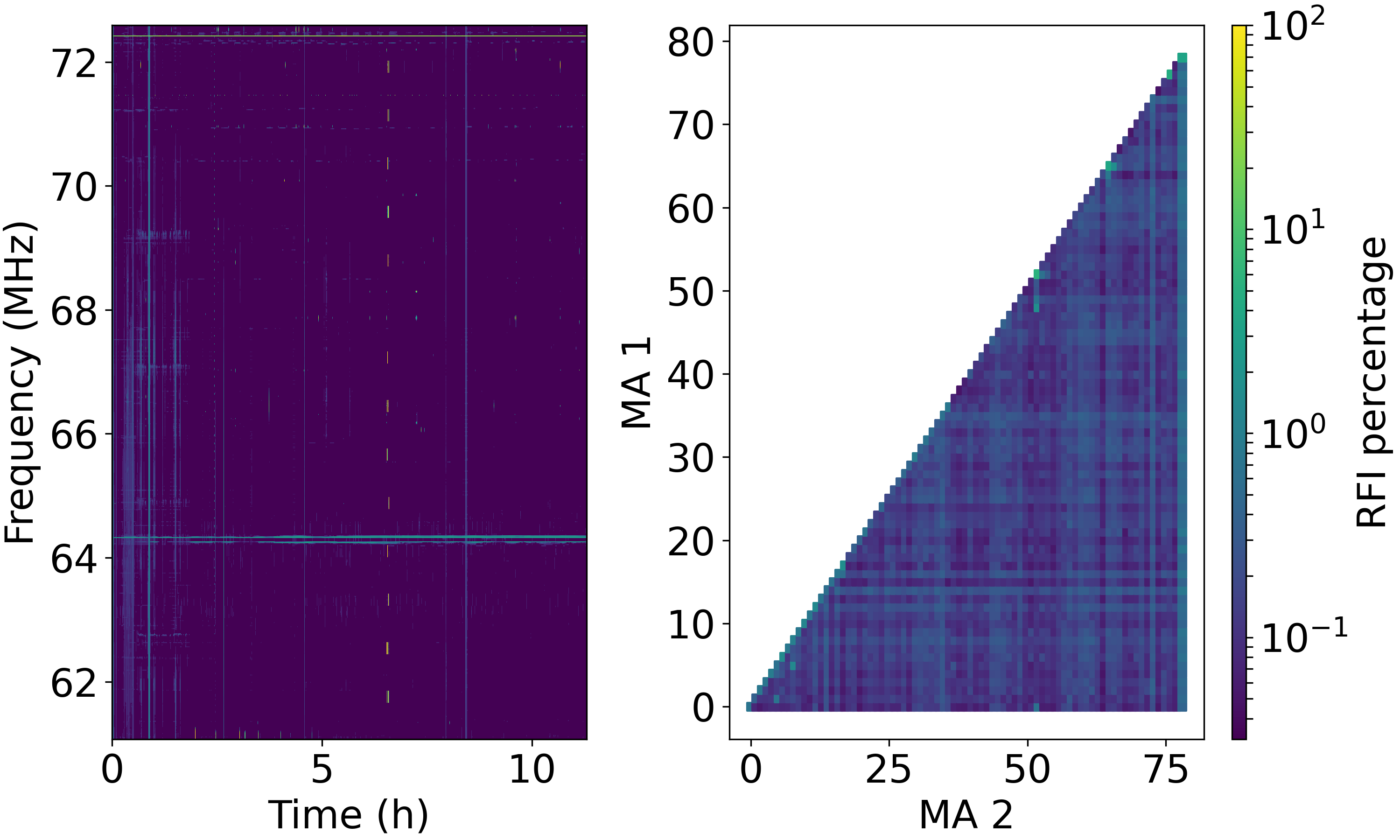}
    \caption{RFI percentage at the L1 data level. The two panels show the flagging percentage as a function of time and frequency (left) and as a function of baseline (right).}
    \label{fig:rfi}
\end{figure}

RFI flagging was performed at different stages of the data analysis pipeline. In Fig. \ref{fig:rfi}, we present the percentage of flagged data by {\tt AOFlagger} until the L1 data level as a function of time, frequency, and baseline.

\FloatBarrier
\section{Calibration solutions}
\subsection{Bandpass solutions}\label{sec:bandpass}
\begin{figure}
        \includegraphics[width=\hsize]{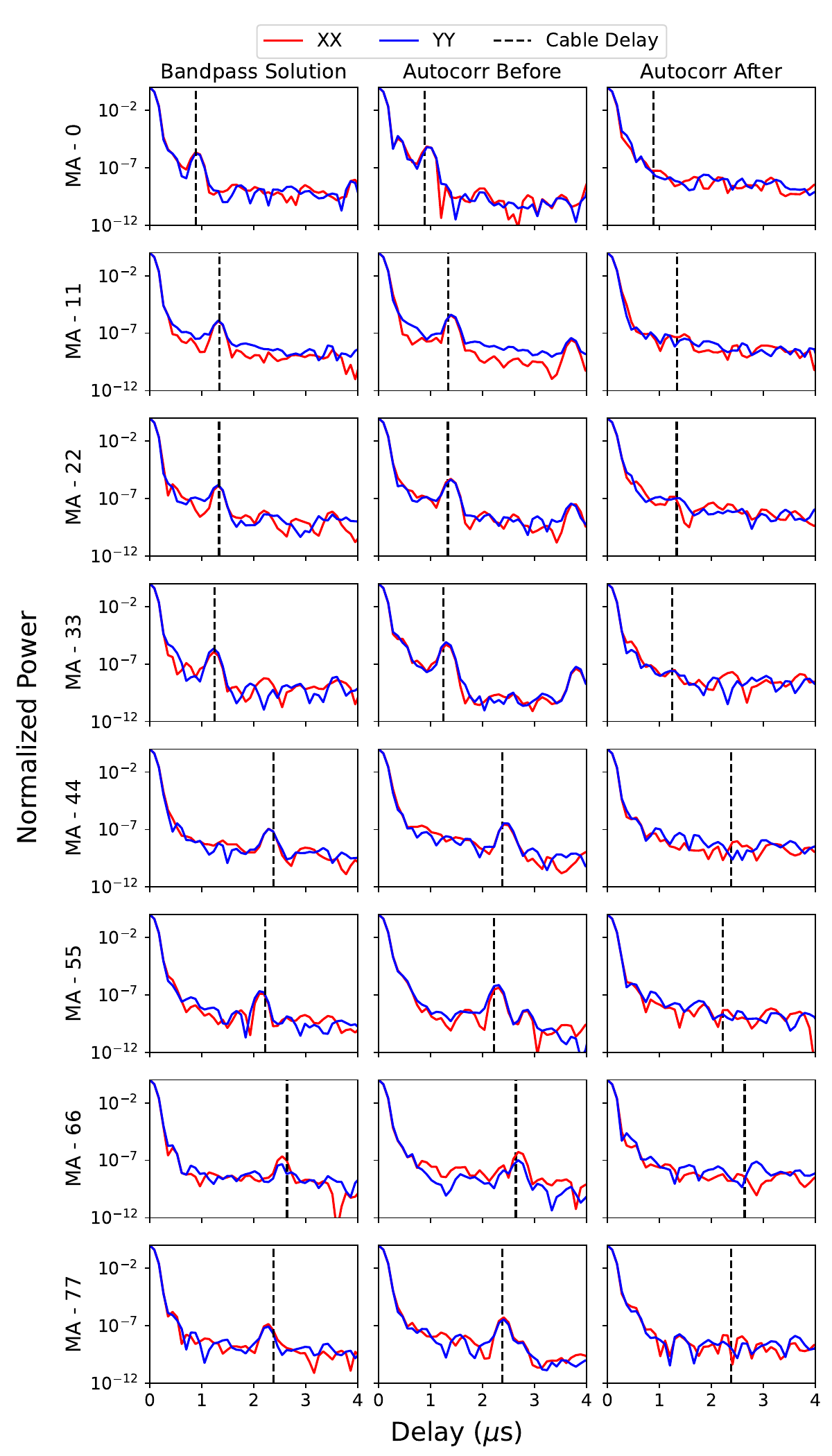}
    \caption{Bandpass calibration results for some example MAs. Left column: Delay power spectra of bandpass solutions obtained from the Cas A observation. Middle column: Delay power spectra of autocorrelations for the NCP observation before bandpass calibration. Right column: Delay power spectra of autocorrelations for the NCP observation after bandpass calibration. The cable lengths for the respective MAs are indicated by vertical dashed lines.}
    \label{fig:bp}
\end{figure}

Bandpass calibration is a crucial step to account for the frequency ripples produced in the data due to cable reflections. In NenuFAR the different MAs have different cable lengths, and the cable reflections are clearly visible in the autocorrelations before bandpass calibration. This is illustrated in Fig. \ref{fig:bp} where the delay power spectra of the autocorrelations of the NCP data before (middle panel) and after (right panel) bandpass calibration are plotted along with the delay power spectra of the bandpass solutions obtained separately from the Cas A observation (left panel), for a few example MAs. The vertical dashed lines indicate the expected delays due to the cable reflections for the MAs calculated from their respective cable lengths. We see a peak near the expected delay in both the bandpass solutions from the Cas A observation and the autocorrelations of the NCP data before bandpass calibration. The peaks are not present in the autocorrelations after the bandpass solutions have been applied to the data.

In the peak-normalized delay power spectra of the autocorrelations before and after bandpass calibration (middle and right panels in Fig. \ref{fig:bp}), we find that even though the peaks due to the cable reflections are accounted for by bandpass calibration, the relative power at high delay values increases after bandpass calibration. To understand the impact of this effect on the power spectra, we performed a test in which the bandpass solutions were smoothed by a Gaussian filter of 2~MHz width, separately on the real and imaginary parts, before applying them to the data. The delay spectra of the autocorrelations with the smoothed bandpass show the peaks at the expected delays corresponding to the cable lengths but have a lower noise floor. However, in the final power spectra, we found negligible differences in the power levels of the autocorrelations with the actual bandpass against the autocorrelations with the smoothed bandpass. This is because the thermal noise level is well above the noise floor of the bandpass gain solutions. For the NCP field, the total flux is not high enough to give a dynamic range, which is needed for these effects to show up in the cylindrical and spherical power spectra.

\subsection{DD gain errors}\label{sec:dd_errors}
\begin{figure}
        \includegraphics[width=\hsize]{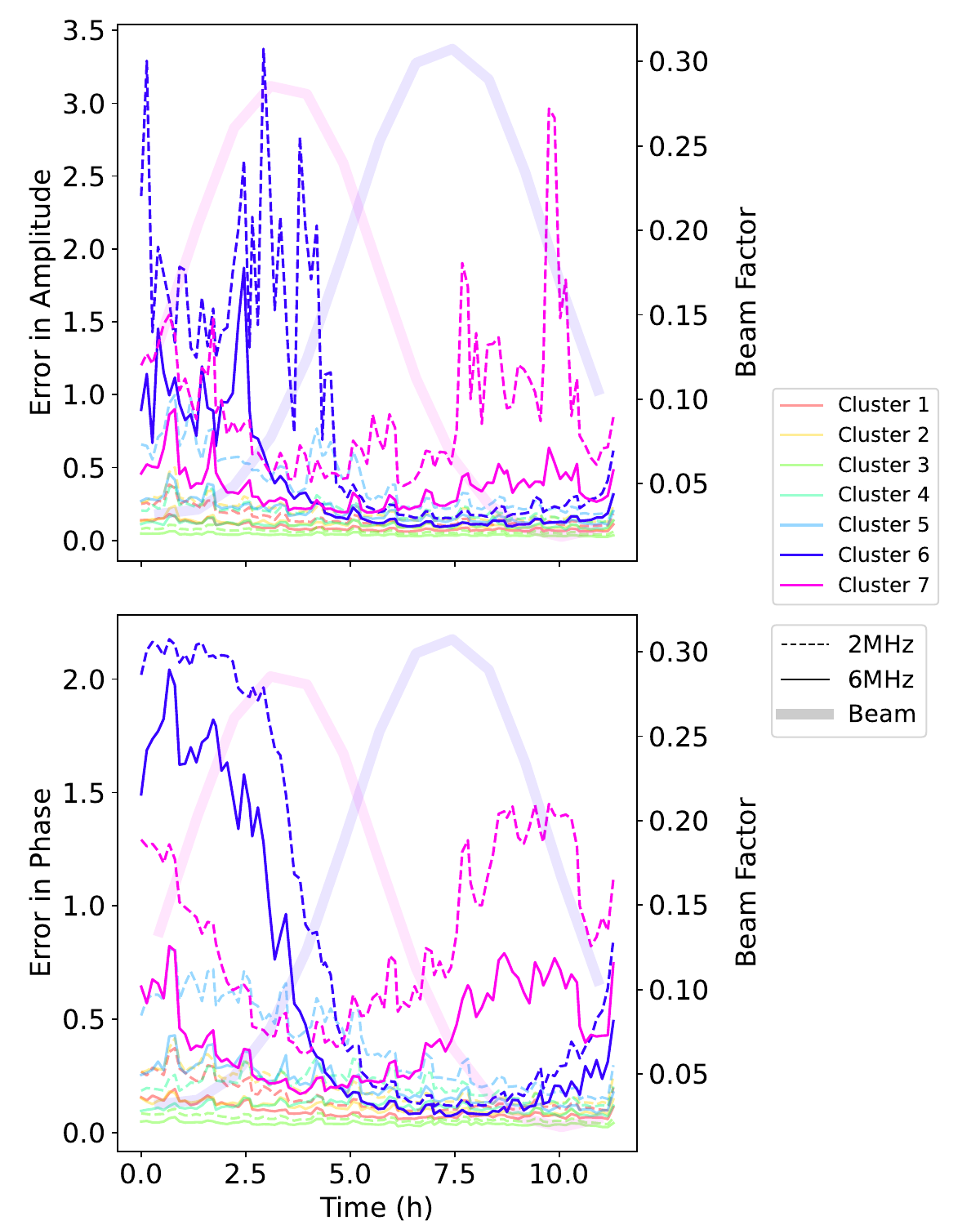}
    \caption{Gain calibration errors at the NCP subtraction step. The top and bottom panels show the errors in the amplitude and phase of the gain solutions, respectively. The dashed lines correspond to the solutions obtained using 2~MHz smoothing while the solid lines correspond to those obtained using 6~MHz smoothing. The thick solid lines show the beam factor and are to be read using the vertical axis on the right-hand side. The different colors refer to the gains in the direction of the different clusters.}
    \label{fig:dd_errors}
\end{figure}

We find that for the NCP subtraction step, a 6~MHz smoothing kernel results in better subtraction of sources compared to a 2~MHz smoothing kernel. Here, we compare the noise in the gain solutions in these two cases. The difference in the gain solutions at successive time steps is used as a proxy for the error in the gain solutions under the assumption that the gains do not change substantially over two solution intervals. Fig. \ref{fig:dd_errors} shows the errors in the amplitude and phase of the DD gain solutions obtained in this manner, by taking the standard deviation of the time-differenced gains across frequencies and stations. Clusters 1 to 7 are arranged in the decreasing order of flux and we see that fainter clusters have higher calibration errors. For each cluster, the calibration errors for a 6~MHz smoothing kernel are smaller compared to a 2~MHz smoothing kernel. The modulation of the errors in the gain solutions toward Clusters 6 and 7 with time is due to the fact that the central lobe of the NenuFAR primary beam is not axisymmetric and is elongated in one direction. As the beam rotates with respect to the sky with time, Cluster 7 overlaps with the beam for the first few hours and then moves out of the beam. So the errors are lower in the first few hours. The opposite is true for Cluster 6 and hence we see the reverse trend in the gain errors. The average primary beam for each data segment at the center of Cluster 6 and Cluster 7 are plotted in Fig. \ref{fig:dd_errors} and the anticorrelation of the gain calibration errors with the primary beam can be seen clearly as a function of time.

\FloatBarrier
\section{A-team subtraction trials}\label{sec:ateamsub_trials}
\begin{figure}
        \includegraphics[width=\hsize]{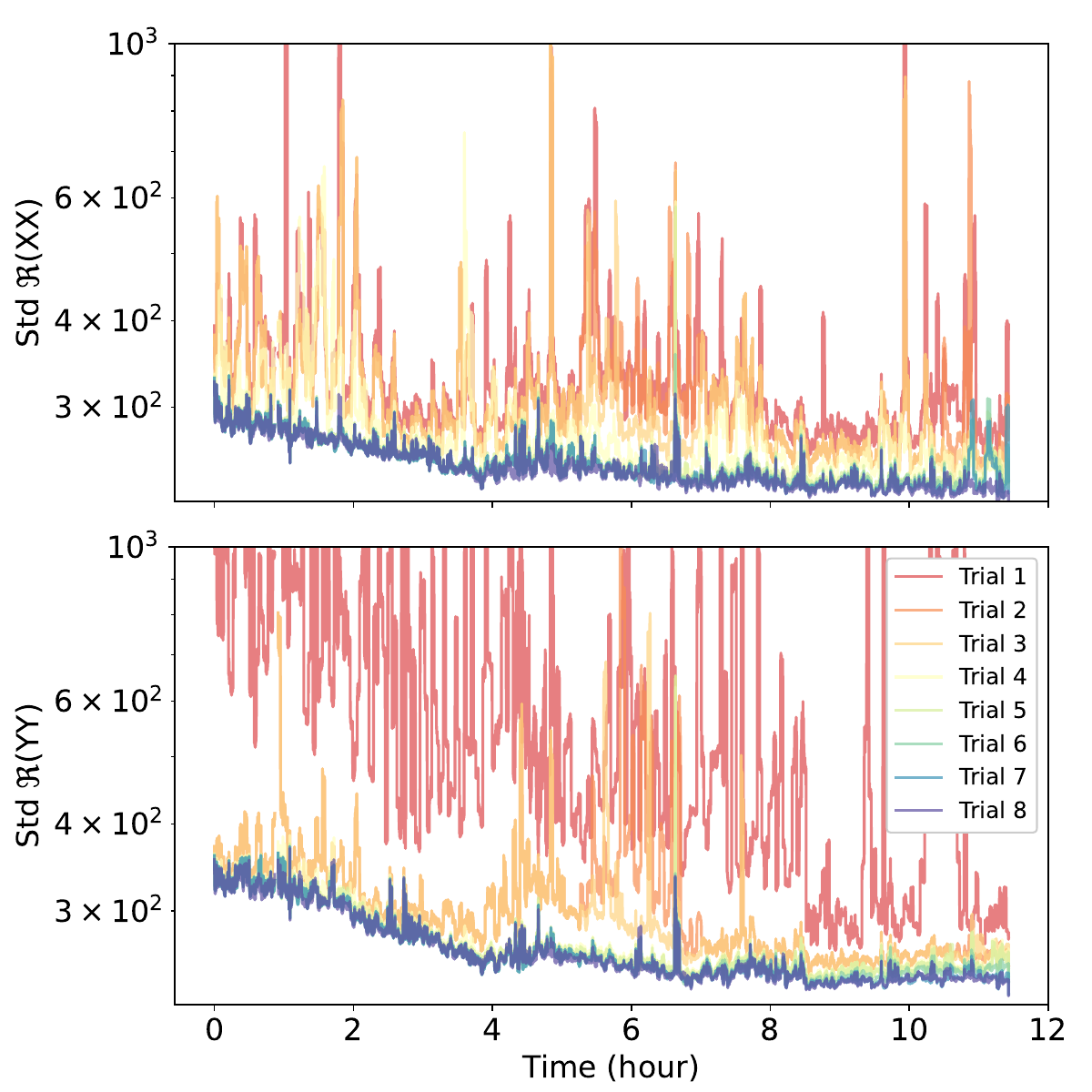}
    \caption{Standard deviation of the real part of the visibilities along frequency and baselines, as a function of time for different calibration trials. At each successive trial, only one setting was changed in order to understand its impact on the calibration. Trial 1 is the result of a {\tt fulljones} calibration with Cyg A and Cas A subtracted when they are at elevations above 10 degrees. The subsequent changes in the other trials are described as follows. Trial 2: MAs 62, 64, 72 flagged, Trial 3: A-team included till 0 degree elevation, Trial 4: Tau A included, Trial 5: MA 6 flagged, Trial 6: MA 18 flagged, Trial 7: Vir A included, Trial 8: Calibration done in the {\tt diagonal} mode.}
    \label{fig:stat_di}
\end{figure}

We varied the number of stations to be flagged, the number of A-team sources to be included, the elevation range for which an A-team is included, and the calibration mode ({\tt fulljones} or {\tt diagonal}) in these trial runs and converged to the optimized calibration parameters by monitoring the calibration solutions, data statistics, and images at the end of each trial. As an example, Fig. \ref{fig:stat_di} shows the standard deviation of the real parts of the XX and YY calibrated visibilities, integrated along all frequencies and baselines, for these different trial runs. We see that it improves progressively as we converge to the final calibration settings. Many more trials were performed with more MAs flagged (MAs 1, 9, 22, 30, 56, 59, and 63), but they did not lead to any further improvement in the solutions, statistics, or images; hence, these stations were not flagged in the final data used for the analysis.

\end{appendix}

\end{document}